\documentclass[seceq,preprint]{ptptex}

\usepackage{amsmath}
\usepackage{amssymb}
\usepackage[dvipdfmx]{graphicx}
\usepackage{subfigure}

\preprintnumber{IU-TH-12}

\title{
Momentum relation and classical limit in the future-not-included complex action theory

}

\author{%
Keiichi \textsc{Nagao}\footnote{E-mail: keiichi.nagao.phys@vc.ibaraki.ac.jp}
and Holger Bech \textsc{Nielsen}\footnote{E-mail: hbech@nbi.dk}
}

\inst{
$^{*)}$Faculty of Education, Ibaraki University, Bunkyo 2-1-1, Mito 310-8512 
Japan \\
{\it and }\\
$^{*),**)}$Niels Bohr Institute, University of Copenhagen, 
17 Blegdamsvej Copenhagen $\phi$, Denmark
}

\abst{%

Studying the time development of the expectation value 
in the future-not-included complex action theory, 
we point out that the momentum relation 
(the relation analogous to $p=\frac{\partial L}{\partial \dot{q}}$), 
which was derived 
via the Feynman path integral 
and was shown to be correct in the future-included theory 
in our previous papers, 
is not valid in the future-not-included theory. 
We provide the correct momentum relation in the future-not-included theory,  
and argue that the future-not-included classical theory 
is described by a certain real action. 
In addition, we provide another way to understand the time development of 
the future-not-included theory by utilizing the future-included theory. 
Furthermore, properly applying the method used in our previous paper 
to the future-not-included theory 
by introducing a formal Lagrangian, 
we derive the correct momentum relation in the future-not-included theory.  
}

\begin{document}

\maketitle

\section{Introduction}

Complex action theory (CAT) is one of the attempts to extend 
quantum theories by allowing their action to be complex. 
CAT has recently been studied 
with the expectation that the imaginary part of the action would give 
some falsifiable predictions\cite{Bled2006,Nielsen:2008cm,Nielsen:2007ak,Nielsen:2005ub}.  
So far, various interesting suggestions have been made for Higgs mass\cite{Nielsen:2007mj}, 
quantum mechanical philosophy\cite{newer1,Vaxjo2009,newer2}, 
some fine-tuning problems\cite{Nielsen2010qq,degenerate}, 
black holes\cite{Nielsen2009hq}, 
de Broglie-Bohm particles and a cut-off in loop diagrams\cite{Bled2010B}. 
Related to CAT, integration contours 
in the complex plane\cite{Garcia:1996np}\cite{Guralnik:2007rx}, 
complex Langevin equations\cite{Pehlevan:2007eq} and 
complexified solution sets\cite{Ferrante:2008yq}\cite{Ferrante:2009gk} 
have also been studied. 

In ref.\cite{Nagao:2010xu}, 
in a system with a non-Hermitian diagonalizable 
bounded Hamiltonian $\hat{H}$, introducing 
a proper inner product\footnote{Similar inner products were also studied 
in refs.\cite{Geyer,Mostafazadeh_CPT_ip_2002,Mostafazadeh_CPT_ip_2003}.}  
and considering the long time development of some states, 
we effectively obtained a Hermitian Hamiltonian. 
We note that $\hat{H}$ is generically non-Hermitian, 
so it does not belong to the class of PT-symmetric non-Hermitian Hamiltonians 
which has been intensively 
studied recently.\cite{Bender:1998ke,Bender:1998gh,Geyer,Mostafazadeh_CPT_ip_2002,
Mostafazadeh_CPT_ip_2003} 
For details of PT-symmetric non-Hermitian Hamiltonians, see 
the reviews\cite{Bender:2005tb,Bender:2007nj,Mostafazadeh:2008pw,Mostafazadeh:2010yx} 
and the references therein. 
In addition, non-Hermitian time-dependent Hamiltonians are studied in 
ref.\cite{Fukuma:2013mx}.
In ref.\cite{Nagao:2011za}, 
introducing various mathematical tools such as a modified set of complex conjugate, 
real and imaginary parts, 
Hermitian conjugates and bras, complex delta function etc., 
we explicitly constructed non-Hermitian operators of coordinate and momentum, 
$\hat{q}_{new}$ and $\hat{p}_{new}$, and the eigenstates of their Hermitian 
conjugates $| q \rangle_{new}$ and $|p \rangle_{new}$  
for complex $q$ and $p$ by utilizing coherent states of harmonic oscillators.  
Indeed, $|q \rangle$, which obeys $\hat{q}|q \rangle = q|q \rangle$, 
is defined only for real $q$, i.e. the eigenvalue of the Hermitian $\hat{q}$, 
so $q$ is not allowed to be complex unless $\hat{q}$ is extended 
to a non-Hermitian operator.  
Only in our complex coordinate formalism can we deal with complex $q$ and $p$. 
This formalism would be a part of proof of consistency 
in using complex $q$ and $p$ in contours of integration for 
WKB (Wentzel-Kramers-Brillouin) approximation, etc. 
in the usual real action theory (RAT). 
Using this formalism in ref.\cite{Nagao:2011is}, 
we explicitly examined the momentum and Hamiltonian 
in the CAT via the Feynman path integral (FPI). 
We studied the time development of some $\xi$-parametrized state, 
which is a solution 
to a kind of eigenvalue problem for a momentum operator.  
Finding the value of $\xi$ that gives 
the largest contribution in FPI, 
we derived the momentum relation $p= m \dot{q}$ and Hamiltonian.

The future-included theory, i.e. the theory including not only 
a past time but also a future time as an integration interval of time, 
was studied in ref.\cite{Bled2006}, 
whose authors introduced 
the future state $| B (T_B) \rangle$ at the final time $T_B=\infty$ 
in addition to the past state $| A (T_A) \rangle$ at the initial time $T_A=-\infty$. 
The states $| A (T_A) \rangle$ and $| B (T_B) \rangle$ 
time-develop according to the non-Hermitian Hamiltonians $\hat{H}$ and 
$\hat{H}_B = \hat{H}^\dag$, respectively. 
The authors of ref.\cite{Bled2006} 
speculated a correspondence of the future-included theory to 
the future-not-included one, 
i.e.$\langle \hat{\cal O} \rangle^{BA} \simeq \langle \hat{\cal O} \rangle^{AA}$, where 
$\langle \hat{\cal O} \rangle^{BA} 
\equiv \frac{ \langle B(t) |  \hat{\cal O}  | A(t) \rangle }{ \langle B(t) | A(t) \rangle }$,  
$\langle \hat{\cal O} \rangle^{AA} \equiv \frac{ \langle A(t) |  \hat{\cal O}  | A(t) \rangle }{ \langle A(t) | A(t) \rangle }$, and $t$ is the present time. 
In the RAT the matrix element 
$\langle \hat{\cal O} \rangle^{BA}$ is called the weak value\cite{AAV}, and 
has been intensively studied. 
For details of the weak value, see 
the reviews\cite{review_wv,AT,AV08,Shikano2011review} 
and the references therein. 
In refs.\cite{Nagao:2012mj,Nagao:2012ye}
we investigated $\langle \hat{\cal O} \rangle^{BA}$ carefully, 
and found that if we regard it as an expectation value, then we 
obtain the Heisenberg equation, Ehrenfest's theorem, and a conserved probability current density. This result strongly suggests that we can regard $\langle \hat{\cal O} \rangle^{BA}$ 
as the expectation value in the future-included theory.  
Furthermore, 
using both the complex coordinate formalism\cite{Nagao:2011za} 
and the automatic hermiticity mechanism\cite{Nagao:2010xu,Nagao:2011za}, 
i.e., a mechanism to obtain a Hermitian 
Hamiltonian after a long time development, 
we obtained a correspondence principle that 
$\langle \hat{\cal O} \rangle^{BA}$ for large $T_B-t$ and large $t-T_A$ is almost 
equivalent to $\langle \hat{\cal O} \rangle_{Q'}^{AA}$ for large $t-T_A$, 
where $Q'$ is a Hermitian operator which is used to define a proper 
inner product. 
Thus the future-included theory is not excluded, although it looks exotic.

As for the momentum relation, in ref.\cite{Nagao:2012mj} we obtained 
$\langle \hat{p}_{new} \rangle^{BA} = m \frac{d}{dt} \langle \hat{q}_{new} \rangle^{BA}$ 
in the case of the future-included theory. 
This is consistent with the momentum relation $p=m \dot{q}$, 
which we derived via FPI in ref.\cite{Nagao:2011is}. 
But how about in the future-not-included theory? 
Here, $\langle \hat{q}_{new} \rangle^{AA}$ and 
$\langle \hat{p}_{new} \rangle^{AA}$ are real, 
if we replace $\hat{q}_{new}$ and $\hat{p}_{new}$ 
with Hermitian $\hat{q}$ and $\hat{p}$ respectively. 
On the other hand, $m\dot{q}$ is complex because $m$ is complex. 
Thus, we encounter a contradiction. 
This is quite in contrast to the case of the future-included theory,  
where $\langle \hat{q}_{new} \rangle^{BA}$ and $\langle \hat{p}_{new} \rangle^{BA}$ 
are complex even if $\hat{q}_{new}$ and $\hat{p}_{new}$ 
are replaced with $\hat{q}$ and $\hat{p}$ respectively, 
so we do not have such a contradiction. 
This fact suggests that the momentum relation $p=m\dot{q}$ is not valid 
in the future-not-included theory.

Thus we are motivated to examine the momentum relation 
in the future-not-included theory. 
In this paper, 
studying the time development of $\langle \hat{\cal O} \rangle^{AA}$, 
we argue that the momentum relation in the future-not-included theory 
is not given by $p=m\dot{q}$ but by another 
expression $p=m_{\text{eff}} \dot{q}$, where 
$m_{\text{eff}}$ is a certain real mass.  
Moreover, since the effect of the anti-Hermitian part of the Hamiltonian 
is suppressed in the classical limit,  
we claim that classical theory in the future-not-included theory is 
described by the real part of the non-Hermitian Hamiltonian, 
or a certain real action $S_{\text{eff}}$. 
In addition, we present another way to understand the time development of 
the future-not-included theory by utilizing the future-included theory. 
Furthermore, we discuss how we can utilize 
the method studied in ref.\cite{Nagao:2011is} 
to obtain the correct momentum relation in the future-not-included theory. 
In the method, we analyze the time development of $\xi$-parametrized 
state in a transition amplitude 
from initial time to final time, where the present time $t$ 
is supposed to be between the initial and final times. 
This is the case for the future-included theory, but not for the future-not-included theory.  
Therefore, to properly apply the method to the future-not-included theory, 
we introduce a formal Lagrangian by 
rewriting the transition amplitude in the future-not-included theory, 
$\langle A(t)| A(t) \rangle$, into an expression such as 
$\langle B(t)| A(t) \rangle$, which is the transition amplitude 
in the future-included theory. 
We argue that using this formal Lagrangian in 
the method we obtain $p=m_{\text{eff}} \dot{q}$, 
the correct momentum relation in the future-not-included theory.

This paper is organized as follows. 
In section 2 we review the complex coordinate formalism proposed 
in ref.~\cite{Nagao:2011za}. 
In section 3, following ref.\cite{Nagao:2011is}, 
we explain the method used to derive the momentum relation $p=m \dot{q}$  
via the Feynman path integral. 
In section 4, based on ref.\cite{Nagao:2012mj}, 
we show that $\langle \hat{\cal O} \rangle^{BA}$ 
behaves as if it were the expectation value of 
some operator $\hat{\cal O}$ in the future-included theory. 
Also, we obtain the relation 
$\langle \hat{p}_{new} \rangle^{BA} = m \frac{d}{dt} \langle \hat{q}_{new} \rangle^{BA}$, which is consistent with the momentum relation 
obtained in ref.\cite{Nagao:2011is}. 
In section 5, studying $\langle O \rangle^{AA}$, 
we obtain the momentum relation in the future-not-included theory, 
$p=m_{\text{eff}} \dot{q}$. 
Moreover, we argue 
that the classical theory is described by a certain real action $S_{\text{eff}}$. 
Furthermore, we provide another way to understand the time development 
of the future-not-included theory by making use of the future-included theory. 
In section 6 we apply the method of ref.\cite{Nagao:2011is} 
to the future-not-included theory properly by introducing the formal Lagrangian, 
and derive the momentum relation in the future-not-included theory, 
which is consistent with that derived in section 5. 
Section 7 is devoted to discussion.

\section{Complex coordinate formalism}\label{sec:review_complex_coordinate}


In this section we briefly review the complex coordinate formalism that we proposed 
in ref.\cite{Nagao:2011za} 
so that we can deal with complex coordinate $q$ and momentum $p$ 
properly not only in the CAT but also in a real action theory (RAT), 
where we encounter them at the saddle point in WKB approximation, etc.

\subsection{Non-Hermitian operators $\hat{q}_{new}$ and $\hat{p}_{new}$, 
and the eigenstates of their Hermitian conjugates $|q \rangle_{new}$ and $|p \rangle_{new}$}

We can construct the non-Hermitian operators  of coordinate and momentum, 
$\hat{q}_{new}$ and $\hat{p}_{new}$, 
and the eigenstates  of their Hermitian conjugates 
$| q \rangle_{new}$ and $| p \rangle_{new}$, such that 
\begin{eqnarray}
&&\hat{q}_{new}^\dag  | q \rangle_{new} =q | q \rangle_{new} , \label{qhatqket=qqket_new} \\
&&\hat{p}_{new}^\dag  | p \rangle_{new} =p | p \rangle_{new} , \label{phatpket=ppket_new} \\
&&[\hat{q}_{new} , \hat{p}_{new}  ] = i \hbar , \label{comqhatphat}
\end{eqnarray}
for complex $q$ and $p$ by formally utilizing two coherent states. 
Our proposal is to replace 
the usual Hermitian operators of coordinate and momentum 
$\hat{q}$, $\hat{p}$, and their eigenstates $|q  \rangle$ and $|p  \rangle$, 
which obey 
$\hat{q} | q \rangle = q| q \rangle$, 
$\hat{p} | p \rangle = p | p \rangle$, 
and $[ \hat{q},  \hat{p} ] = i \hbar$ for real $q$ and $p$, 
with $\hat{q}_{new}^\dag$, $\hat{p}_{new}^\dag$,  
$|q \rangle_{new}$ and $|p \rangle_{new}$. 
The explicit expressions for $\hat{q}_{new}$, $\hat{p}_{new}$, 
$| q \rangle_{new}$ and $| p \rangle_{new}$
are given by\footnote{For simplicity we have replaced 
the parameters $m\omega$ and $m'\omega'$ used in ref.\cite{Nagao:2011za} with 
$\frac{1}{\epsilon}$ and $\epsilon'$.}
\begin{eqnarray}
&&\hat{q}_{new} \equiv 
\frac{1}{ \sqrt{1 - \epsilon \epsilon' }  } \left( \hat{q} - i \epsilon \hat{p}  \right), \label{def_qhat_new} \\
&&\hat{p}_{new} \equiv 
\frac{1}{ \sqrt{1 - \epsilon \epsilon' }  }  \left( \hat{p} + i \epsilon' \hat{q} \right) , \label{def_phat_new} \\
&&| q \rangle_{new} 
\equiv 
\left( \frac{1 - \epsilon \epsilon' }{4\pi \hbar \epsilon } \right)^{\frac{1}{4}} 
e^{- \frac{1}{4\hbar\epsilon  } \left( 1 - \epsilon \epsilon' \right) {q}^2 }
| \sqrt{ \frac{1 - \epsilon \epsilon'}{2\hbar \epsilon} } q \rangle_{coh} ,  \\
&&| p \rangle_{new} 
\equiv 
\left( \frac{1 - \epsilon \epsilon' }{4\pi \hbar \epsilon' } \right)^{\frac{1}{4}} 
e^{ -\frac{1}{4 \hbar \epsilon'}  \left( 1 - \epsilon \epsilon' \right)  p^2 }
| i \sqrt{ \frac{ 1 - \epsilon \epsilon'}{2\hbar \epsilon'} }  p \rangle_{coh'} , \label{defpketnew}
\end{eqnarray}
where $| \lambda \rangle_{coh}$ is a coherent state parametrized with a complex parameter $\lambda$ defined up to a normalization factor by 
$| \lambda \rangle_{coh} 
\equiv  
e^{\lambda a^\dag} | 0 \rangle 
= \sum_{n=0}^{\infty} \frac{\lambda^n}{\sqrt{n!}} | n \rangle$, 
and this satisfies the relation 
$a | \lambda \rangle_{coh} = \lambda | \lambda \rangle_{coh}$. 
Here, 
$a = \sqrt{ \frac{1}{2\hbar \epsilon  }}  \left( \hat{q} + i  \epsilon \hat{p} \right)$ and 
$a^\dag = \sqrt{ \frac{1}{2\hbar \epsilon  }}\left( \hat{q} - i  \epsilon \hat{p} \right)$
are annihilation and creation operators. 
In eq.(\ref{defpketnew}), 
$| \lambda \rangle_{coh'} \equiv e^{\lambda {a'}^\dag} | 0 \rangle$, 
where ${a'}^\dag$  is given by 
${a'}^\dag = \sqrt{ \frac{\epsilon'}{2\hbar}}
\left( \hat{q} - i \frac{ \hat{p}}{\epsilon'}  \right) \label{creation'}$, 
is another coherent state defined similarly. 
Before seeing the properties of $\hat{q}_{new}$, $\hat{p}_{new}$, 
$| q \rangle_{new}$, and $| p \rangle_{new}$, 
we define a delta function of complex parameters in the next subsection.

\subsection{The delta function}\label{subsec:deltafunc}

\begin{figure}[htb]
\begin{center}
\includegraphics[height=10cm]{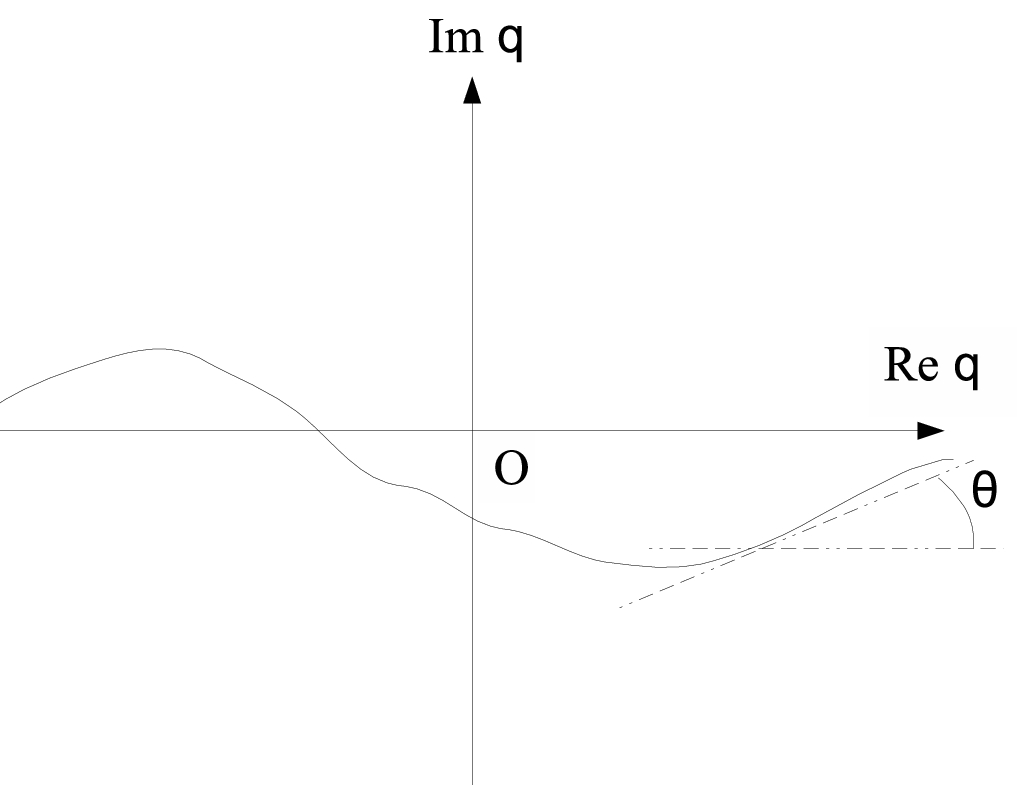}
\end{center}
\caption{An example permitted path $C$}
\label{fig:contour}
\end{figure}

\begin{figure}[htb]
\begin{center}
\includegraphics[height=10cm]{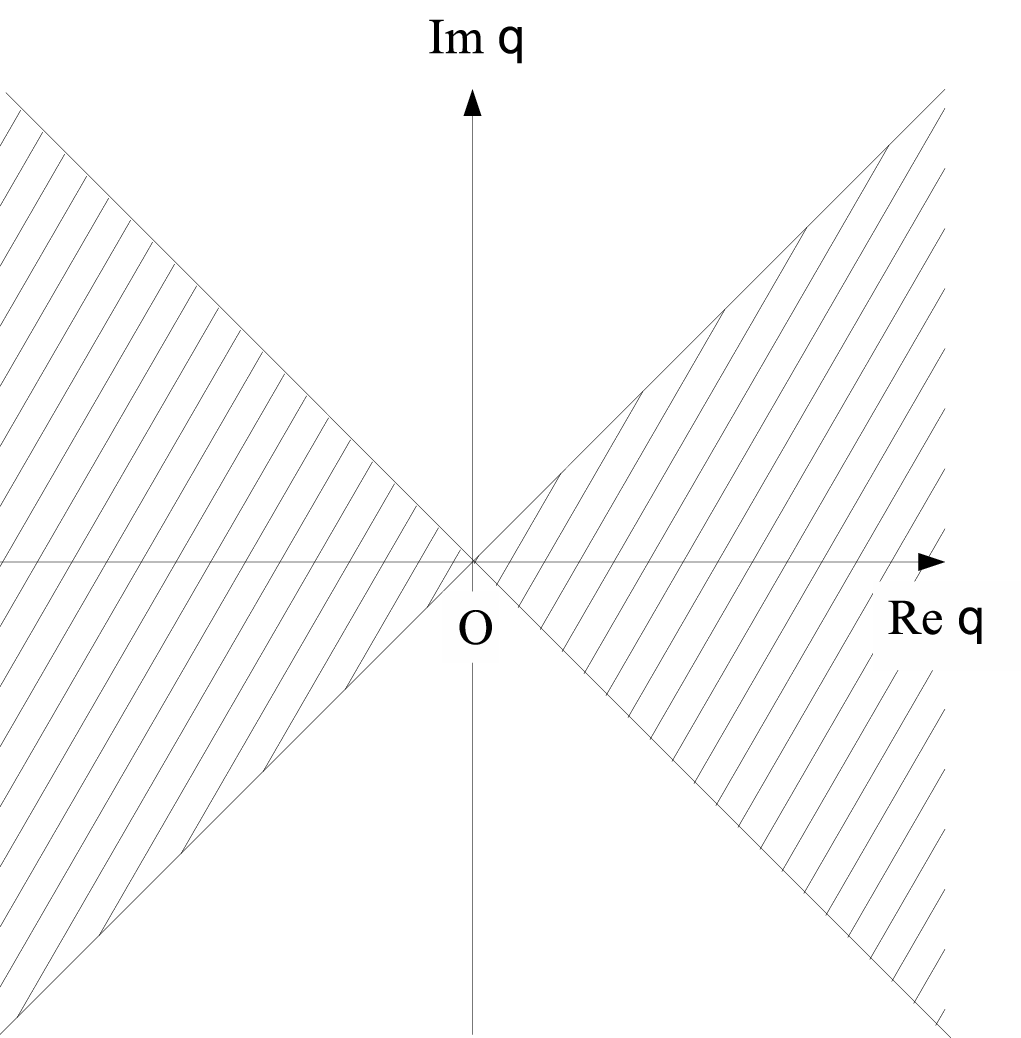}
\end{center}
\caption{Domain of the delta function}
\label{fig:delta_function}
\end{figure}

We define ${\cal D}$ as 
a class of distributions depending on one complex variable $q \in \mathbf{C}$. 
Using a function $g:{\mathbf C} \rightarrow {\mathbf C}$ as 
a distribution\footnote{We recently noticed that another complex distribution 
was introduced in ref.\cite{Nakanishi}. 
It is different from ours in the following points:
the complex distribution in ref.\cite{Nakanishi}, where $g(q)$ is supposed to have poles, 
is not well defined by $g(q)$ alone, but needs the indication of which side of the 
poles the path $C$ passes through. On the other hand, in our complex distribution 
we assume not the presence of poles of $g(q)$ but $f$ not being a bounded entire 
function.} 
in the class ${\cal D}$, 
we define the following functional $G$ 
\begin{equation}
G[f] = \int_C f(q) g(q) dq  \label{mappingG}
\end{equation}
for any analytical function $f:{\mathbf C} \rightarrow {\mathbf C}$ 
with convergence requirements such that $f \rightarrow 0$ for $q \rightarrow \pm \infty$. 
The functional $G$ is a linear mapping from the function $f$ to a complex number. 
Since the simulated function $g$ is supposed to be analytical in $q$, 
the path $C$, which is chosen 
to run from $-\infty$ to $\infty$ in the complex plane, 
can be deformed freely and so it is not relevant. 
As an example of such a distribution we could think of the delta function
and approximate it by 
the smeared delta function defined for complex $q$ by 
\begin{equation}
g(q) = \delta_c^\epsilon(q) 
\equiv \sqrt{\frac{1}{4 \pi \epsilon}} e^{-\frac{q^2}{4\epsilon}} , \label{delta_c_epsilon(q)}
\end{equation}
where $\epsilon$ is a finite small positive real number. 
For the limit of $\epsilon \rightarrow 0$, $g(q)$ converges in the distribution sense 
for complex $q$ obeying the condition 
\begin{equation}
L(q) 
\equiv \left( \text{Re}(q) \right)^2 - \left( \text{Im}(q) \right)^2 
>0 . \label{cond_of_q_for_delta} 
\end{equation}
For any analytical test function $f(q)$\footnote{Because of the Liouville theorem, if $f$ is 
a bounded entire function, 
$f$ is constant. So we are considering $f$ as an unbounded entire function or a function 
that is not entire but is holomorphic at least in the region on which the path runs.} 
and any complex $q_0$, this $\delta_c^\epsilon(q)$ satisfies 
\begin{equation}
\int_C f(q) \delta_c^\epsilon(q-q_0) dq = f(q_0) ,
\end{equation}
as long as we choose the path $C$ 
such that it runs from $-\infty$ to $\infty$ in the complex plane 
and at any $q$ its tangent line and a horizontal line 
form an angle $\theta$ whose absolute value  
is within $\frac{\pi}{4}$ to satisfy the inequality (\ref{cond_of_q_for_delta}). 
An example permitted path is shown in Fig.\ref{fig:contour}, 
and the domain of the delta function is shown in Fig.\ref{fig:delta_function}. 
%


Next, we extend the delta function to complex $\epsilon$, 
and consider 
\begin{eqnarray}
\delta_c^{\epsilon}(aq) 
&=& \sqrt{\frac{1}{4 \pi \epsilon}} e^{-\frac{1}{4\epsilon} a^2 q^2} 
\label{deltaepsilonaq}
\end{eqnarray}
for non-zero complex $a$. 
We express $\epsilon$, $q$, and $a$ as 
$\epsilon = r_\epsilon e^{i\theta_\epsilon}$, 
$q= r e^{i\theta}$, and $a = r_a e^{i\theta_a}$. 
The convergence condition of $\delta_c^{\epsilon}(aq)$: 
$\text{Re} \left( \frac{a^2 q^2}{\epsilon} \right) > 0$ 
is expressed as 
\begin{eqnarray}
&&-\frac{\pi}{4} + \frac{1}{2} ( \theta_{\epsilon} - 2\theta_a ) 
< \theta < \frac{\pi}{4} + \frac{1}{2} ( \theta_{\epsilon}  - 2 \theta_a ), 
\label{cond1forqa} \\
&&\frac{3}{4} \pi + \frac{1}{2}( \theta_{\epsilon} - 2 \theta_a ) 
< \theta < \frac{5}{4}\pi + \frac{1}{2} ( \theta_{\epsilon} - 2 \theta_a ) . 
\label{cond2forqa}
\end{eqnarray}
For $q$, $\epsilon$, and $a$ such that eqs.(\ref{cond1forqa})(\ref{cond2forqa}) 
are satisfied, $\delta_c^{\epsilon}(aq)$ behaves well as a delta function 
of $aq$, and we obtain the relation 
\begin{equation}
\delta_c^{\epsilon}(aq) = 
\frac{\text{sign}(\text{Re} a)}{ a } \delta_c^{\frac{\epsilon}{a^2}}(q) , 
\label{scaling_deltafunction} 
\end{equation}
where we have introduced an expression 
\begin{eqnarray}
\text{sign}(\text{Re} a) 
\equiv 
\left\{ 
\begin{array}{cc}
1  & \text{for} ~\text{Re}a > 0 , \\
-1 & \text{for} ~\text{Re}a < 0 . \\ 
\end{array}
\right.
\end{eqnarray}
%

\subsection{New devices to handle complex parameters}\label{newdevices}

To keep the analyticity in dynamical variables of FPI such as 
$q$ and $p$, 
we define a modified set of a complex conjugate, 
real and imaginary parts, bras and Hermitian conjugates.

\subsubsection{Modified complex conjugate $*_{ \{ \} }$}

We define a modified complex conjugate for a function of $n$ parameters 
$f( \{a_i \}_{i=1, \ldots, n} )$ by 
\begin{equation}
f(\{a_i \}_{i=1, \ldots, n} )^{*_{\{a_i | i \in A \}} } = f^*( \{a_i \}_{i \in A}  ,  \{a_i ^*\}_{i \not\in A} ) , 
\end{equation}
where $A$ denotes the set of indices attached to the parameters 
in which we keep the analyticity, 
and $*$ on $f$ acts on the coefficients included in $f$. 
For example, the complex conjugate 
$*_{q,p}$ 
of a function $f(q,p)=a q^2 + b p^2$ is  
written as $f(q,p)^{*_{q,p}} = a^* q^2 + b^* p^2$.
The analyticity is kept in both $q$ and $p$. 
For simplicity we express the modified complex conjugate as $*_{ \{  \} }$, 
where $\{ \}$ is a symbolic expression for a set of parameters 
in which we keep the analyticity.

\subsubsection{Modified real and imaginary parts $\text{Re}_{\{ \}}$, $\text{Im}_{\{ \}}$ }

We define the modified real and imaginary parts by using $*_{ \{  \} }$.
We decompose some complex function $f$ as 
\begin{equation}
f= \text{Re}_{\{ \}} f + i \text{Im}_{\{ \}} f ,
\end{equation}
where $\text{Re}_{\{ \}} f$ and $\text{Im}_{\{ \}} f$ are 
the ``$\{ \}$-real" and ``$\{ \}$-imaginary" parts of $f$ defined by 
\begin{eqnarray}
&&\text{Re}_{\{ \}} f \equiv \frac{ f + f^{*_{\{ \}}}   }{2} , \label{{}-real} \\
&&\text{Im}_{\{ \}} f \equiv \frac{ f - f^{*_{\{ \}}}   }{2i} . \label{{}-imaginary}
\end{eqnarray}
For example, for $f=k q^2$, the $q$-real and $q$-imaginary parts of $f$ 
are expressed as 
$\text{Re}_{q} f = \text{Re} ( k ) q^2$ and 
$\text{Im}_{q} f = \text{Im} ( k ) q^2$, respectively. 
In particular, if $f$ satisfies $f^{*_{\{ \}}} =f$, 
we say $f$ is $\{ \}$-real, while if $f$ obeys $f^{*_{\{ \}}} =-f$, 
we call $f$ purely $\{ \}$-imaginary. 


\subsubsection{Modified bras ${}_m \langle ~|$ and ${}_{ \{ \} } \langle ~|$, 
and modified Hermitian conjugate $\dag_{ \{ \} }$ }

For some state $| \lambda \rangle$ with some complex parameter $\lambda$, 
we define a modified bra ${}_m\langle \lambda |$ by 
\begin{equation} 
{}_m\langle \lambda | \equiv \langle \lambda^* |  \label{modified_bra_anti-linear}
\end{equation}
so that it preserves the analyticity in $\lambda$. 
In the special case of $\lambda$ being real it becomes a normal bra. 
In addition we define a slightly generalized modified bra 
${}_{\{\}}\langle ~|$ and a modified Hermitian conjugate $\dag_{ \{ \} }$ of a ket. 
For example, ${}_{u,v}\langle u | = {}_u \langle u | = {}_m\langle u |$, 
$( | u \rangle )^{\dag_{u, v}}  =( | u \rangle )^{\dag_{u}}  = {}_m \langle u |$. 
We express the Hermitian conjugate $\dag_{ \{ \} }$ of a ket symbolically as 
$( |  ~\rangle )^{\dag_{\{ \} }} = {}_{\{ \}}\langle  ~|$. 
Also, we write the Hermitian conjugate $\dag_{ \{ \} }$ of a bra as 
$( {}_{\{ \}}\langle  ~| )^{\dag_{\{ \} }} =  |  ~\rangle$. 
Hence, for a matrix element we have the relation 
${}_{\{\}}\langle u | A | v \rangle^{*_{ \{ \} }} = 
{}_{ \{ \} } \langle v | A^\dag | u \rangle$.

\subsection{Properties of $\hat{q}_{new}$, $\hat{p}_{new}$, $|q \rangle_{new}$ and $|p \rangle_{new}$, 
and a theorem for matrix elements} \label{prop_qnewpnew}

The states $| q \rangle_{new}$ and $| p \rangle_{new}$ are normalized 
so that they satisfy 
the following relations: 
\begin{eqnarray}
{}_m\langle_{new}~ q' | q \rangle_{new} 
&=& \delta_c^{\epsilon_1} ( q' - q ) , \label{m_q'branew_qketnew}\\ 
{}_m\langle_{new}~ p' | p \rangle_{new} 
&=& \delta_c^{\epsilon'_1} ( p' - p ) , \label{m_p'branew_pketnew}
\end{eqnarray}
where $\epsilon_1 = \frac{\hbar \epsilon }{ 1 -  \epsilon \epsilon' }$ 
and $\epsilon'_1 = \frac{\hbar \epsilon'}{1 - \epsilon \epsilon' }$. 
We take $\epsilon$ and $\epsilon'$ sufficiently small, 
for which the delta functions converge 
for complex $q$, $q'$, $p$, and $p'$ satisfying the conditions 
$L(q-q') > 0$ and $L(p-p') > 0$, where $L$ is given in eq.(\ref{cond_of_q_for_delta}). 
These conditions are satisfied only when $q$ and $q'$ or $p$ and $p'$ are 
on the same paths respectively.  
Eqs.(\ref{m_q'branew_qketnew})(\ref{m_p'branew_pketnew}) represent 
the orthogonality relations for $| q \rangle_{new}$ and $| p \rangle_{new}$, 
and we have the following relations for complex $q$ and $p$: 
\begin{eqnarray}
&&\int_C dq | q \rangle_{new} ~{}_m \langle_{new} q |  = 1 , \label{completion_complexq_ket2} \\
&&\int_C dp | p \rangle_{new} ~{}_m \langle_{new} p |  = 1 , \label{completion_complexp_ket2} \\
&&\hat{p}_{new}^\dag | q \rangle_{new} 
=i \hbar \frac{\partial}{\partial q} | q \rangle_{new}, \label{phatnewqketnew2} \\
&&\hat{q}_{new}^\dag | p \rangle_{new} 
= \frac{\hbar}{i} \frac{\partial}{\partial p} | p \rangle_{new}, \label{qhatnewpketnew2} \\
&&{}_m\langle_{new}~ q | p \rangle_{new} 
= \frac{1}{\sqrt{2 \pi \hbar}} \exp\left(\frac{i}{\hbar}p q \right). 
\label{basis_fourier_transf2}  
\end{eqnarray}
Thus, $\hat{q}_{new}^\dag$, $\hat{p}_{new}^\dag$, 
$| q \rangle_{new}$ and $| p \rangle_{new}$ 
with complex $q$ and $p$ 
obey the same relations as $\hat{q}$, $\hat{p}$, $| q \rangle$, 
and $| p \rangle$ with real $q$ and $p$. 
In the limits of $\epsilon \rightarrow 0$ and $\epsilon' \rightarrow 0$ 
$\delta_c^{\epsilon_1} ( q' - q )$, $\delta_c^{\epsilon'_1} ( p' - p )$, and 
$\exp\left( \frac{i}{\hbar}p q \right)$ 
in eqs.(\ref{m_q'branew_qketnew})(\ref{m_p'branew_pketnew})(\ref{basis_fourier_transf2}) 
are well defined as distributions of the type ${\cal D}$. 
For real $q'$ and $p'$, $| q' \rangle_{new}$ and $| p' \rangle_{new}$ become 
$| q' \rangle$ and $| p' \rangle$ respectively; 
also, $\hat{q}_{new}^\dag$ and $\hat{p}_{new}^\dag$ behave like  
$\hat{q}$ and $\hat{p}$ respectively. 
In addition, we have the following theorem. 

\vspace{0.5cm}
\noindent
Theorem 1: 
The matrix element 
${}_m\langle_{new}~ q' ~\text{or}~ p'|
{\cal O}( \hat{q}_{new}, \hat{q}_{new}^\dag, \hat{p}_{new}, \hat{p}_{new}^\dag) | q'' ~\text{or}~ p'' \rangle_{new}$, 
where ${\cal O}$ is a Taylor-expandable function, 
can be evaluated as if inside ${\cal O}$ 
we had the hermiticity conditions 
$\hat{q}_{new} \simeq \hat{q}_{new}^\dag \simeq \hat{q}$ and 
$\hat{p}_{new} \simeq  \hat{p}_{new}^\dag \simeq \hat{p}$ 
for $q'$, $q''$, $p'$, $p''$ such that the resulting quantities  
are well defined in the sense of distribution.  
\vspace{0.5cm}

\noindent
This theorem is understood by noticing that 
such a matrix element 
can be expressed as the summation of the products of factors made 
of $q'$, $p'$, $q''$, $p''$ or their differential operators and distributions. 
Thus, we do not have to worry about 
the anti-Hermitian terms in $\hat{q}_{new}$, $\hat{q}_{new}^\dag$, $\hat{p}_{new}$ 
and $\hat{p}_{new}^\dag$, 
provided that we are satisfied with the result in the distribution sense.

\section{Deriving the momentum relation 
via Feynman path integral}  \label{sec:review_xi-state_argument}

We briefly explain how we derived the momentum relation 
in ref.\cite{Nagao:2011is}.

\subsection{The derivation of the momentum relation and the Hamiltonian}
\label{review_derivation_p_and_H}

The FPI in the CAT is described with the following Lagrangian 
-- a typical example for a system with a single degree of freedom --: 
\begin{equation}
L(q(t), \dot{q}(t))=\frac{1}{2}m \dot{q}^2- V(q) , \label{lagrangian}
\end{equation}
where $V(q)=\sum_{n=2}^\infty b_n q^n$ is a potential term. 
For our later convenience we decompose $V$ and $L$ as 
$V=V_R + iV_I$ and $L = L_R + i L_I$, 
where $V_R$, $V_I$, $L_R$ and $L_I$ are given by 
\begin{eqnarray}
&&V_R \equiv \text{Re}_q (V) =\sum_{n=2}^\infty \text{Re} b_n ~q^n , \label{V_R} \\
&&V_I \equiv \text{Im}_q (V) =\sum_{n=2}^\infty \text{Im} b_n ~q^n , \label{V_I} \\
&&L_R \equiv \text{Re}_q (L) = \frac{1}{2}m_R \dot{q}^2 - V_R(q) , \label{L_R} \\
&&L_I \equiv \text{Im}_q (L) = \frac{1}{2}m_I \dot{q}^2 - V_I(q) . \label{L_I}
\end{eqnarray}
Here, $\text{Re}_q$ and $\text{Im}_q$ are as introduced 
in eqs.(\ref{{}-real})(\ref{{}-imaginary}), and 
we have decomposed $m$ into its real and imaginary parts as $m=m_R + i m_I$.

We consider the functional integral  
$\int_C e^{\frac{i}{\hbar} \int L(q, \dot{q}) dt } {\cal D} q$ 
by discretizing the time direction and writing $\dot{q}$ as 
$\dot{q} = \frac{ q(t+ dt) - q(t) }{dt}$, 
where $dt$ is assumed to be a small quantity. 
Since we use the Schr\"{o}dinger representation for wave functions, 
to avoid the confusion with the Heisenberg representation 
we introduce the notations $q_t \equiv q(t)$ and  
$q_{t+dt} \equiv q(t+dt)$, which we regard as independent variables. 
We suppose that the asymptotic values of dynamical variables such as $q$ and $p$ are 
on the real axis, while parameters such as $m$ and $b_n$ are complex in general. 
The path $C$ denotes an arbitrary path running from $-\infty$ to $\infty$ in the complex plane, 
and we can deform it as long as the integrand keeps the analyticity in $q$ and $p$. 
To prevent the kinetic term in the integrand 
from blowing up for $\dot{q}\rightarrow \pm\infty$ along the real axis 
we impose the condition $m_I \geq 0$ on $m$.

In FPI the time development of some wave function 
${}_m \langle_{new}~ q_t | \psi(t) \rangle$ at some time $t$ to $t+dt$ is described by  
\begin{equation}
{}_m \langle_{new}~ q_{t+dt} | \psi(t+dt) \rangle 
= \frac{1}{\alpha(dt)}
\int_C e^{\frac{i}{\hbar} dt L(q,\dot{q})}
{}_m \langle_{new}~ q_{t} | \psi(t) \rangle d q_t , \label{time_dev_qbraAket}
\end{equation}
where $L(q, \dot{q})$ is given by eq.(\ref{lagrangian}), and 
$C$ is an arbitrary path running from $-\infty$ to $\infty$ in the complex plane. 
In addition, $\alpha(dt)$ is a $dt$-dependent normalization factor, 
which is properly fixed later. 
In ref.\cite{Nagao:2011is}, to derive the momentum relation 
$p=\frac{\partial L}{\partial \dot{q}}$, 
we considered some wave function ${}_m \langle_{new}~ q_t | \xi \rangle$ 
that obeys 
\begin{eqnarray}
{}_m \langle_{new}~ q_t | \hat{p}_{new} | \xi \rangle 
&=&
\frac{\hbar}{i} \frac{\partial}{\partial q_t}~{}_m \langle_{new}~ q_t | \xi \rangle \nonumber \\ 
&=&
\frac{\partial L}{\partial \dot{q}}\left( q_t, \frac{\xi-q_t}{dt} \right) 
{}_m \langle_{new}~ q_t | \xi \rangle , \label{phat_p_xi2}
\end{eqnarray}
where $\xi$ is any number. 
Since the set $\left\{ |\xi \rangle \right\}$ is an approximately reasonable basis which has 
roughly completeness 
$1 \simeq \int_C d\xi | \xi \rangle ~{}_m \langle \text{anti} ~\xi |$ 
and orthogonality 
${}_m \langle \text{anti} ~\xi | \xi' \rangle \simeq \delta_c(\xi -\xi')$, 
where ${}_m \langle \text{anti} ~\xi |$ is a dual basis of $| \xi \rangle$, 
we can expand the wave function ${}_m \langle_{new}~ q_t | \psi(t) \rangle$ into 
a linear combination of ${}_m \langle_{new}~ q_t | \xi \rangle$ as 
\begin{eqnarray}
{}_m \langle_{new}~ q_t | \psi(t) \rangle 
&\simeq& \int_C d\xi  ~{}_m \langle_{new}~ q_t | \xi \rangle ~{}_m \langle \text{anti} ~\xi | \psi(t) \rangle \nonumber \\
&=& 
\int_C d\xi  ~{}_m \langle_{new}~ q_t | \psi(t) \rangle|_\xi . 
\end{eqnarray}
Then, solving eq.(\ref{phat_p_xi2}), we obtain 
\begin{eqnarray}
{}_m \langle_{new}~ q_{t+dt} | \psi(t+dt) \rangle|_\xi 
&=&
\frac{1}{\alpha(dt)} \sqrt{\frac{2\pi \hbar dt}{m}}  ~{}_m \langle \text{anti} ~\xi | \psi(t) \rangle 
\exp\left[  \frac{i m}{2\hbar dt}  (q_{t+dt}^2 - \xi^2 ) \right] \nonumber \\
&&\times
\left\{
\delta_c(\xi - q_{t+dt}) 
- \sum_{n=2}  \left(  \frac{\hbar dt}{m} \right)^{n} (-i)^n 
\frac{i dt}{\hbar}  b_n 
\frac{ \partial^n \delta_c(\xi - q_{t+dt} ) }{  \partial \xi^n }  \right\} . \nonumber \\ 
\label{time_dev_qbraAket_k=0general}
\end{eqnarray}
Since ${}_m \langle_{new}~ q_{t+dt} | \psi(t+dt) \rangle|_\xi$ is equal to 
the linear combination of $\delta_c (q_{t+dt} -\xi)$ and its derivative, 
only the component with $\xi=q_{t+dt}$ contributes to 
${}_m \langle_{new}~ q_{t+dt} | \psi(t+dt) \rangle$. 
Thus, we have obtained the momentum relation in the sense of eq.(\ref{phat_p_xi2}): 
\begin{equation}
p=\frac{\partial L}{\partial \dot{q}} = m \dot{q}. \label{def_p_delLdelqdot} 
\end{equation}

Furthermore, we can estimate the right-hand side of eq.(\ref{time_dev_qbraAket}) 
explicitly as follows:  
\begin{eqnarray}
{}_m \langle_{new}~ q_{t+dt} | \psi(t+dt) \rangle 
&=&
\frac{1}{\alpha(dt)} \int_{C'} d\xi
\int_{C} d q_t e^{\frac{i}{\hbar} dt L(q,\dot{q})}
~{}_m \langle_{new}~ q_t | \xi \rangle ~{}_m \langle \text{anti} ~\xi | \psi(t) \rangle
\nonumber \\
&\simeq& 
~{}_m \langle_{new}~  q_{t+dt} |  \exp \left( -\frac{i}{\hbar} \hat{H} dt 
\right)  |\psi(t) \rangle , \label{time_dev_qbraAket_k=0_3}
\end{eqnarray}
where we have taken $\alpha(dt)= \sqrt{\frac{2\pi i \hbar dt}{m}}$ so that both sides of 
eq.(\ref{time_dev_qbraAket_k=0_3}) 
correspond to each other in the vanishing limit of $dt$, and $\hat{H}$ is given by  
\begin{equation}
\hat{H}
=H(\hat{q}_{new} , \hat{p}_{new} ) 
= \frac{1}{2m} (\hat{p}_{new})^2 + V(\hat{q}_{new}) . \label{expHhat}
\end{equation}
Then eq.(\ref{time_dev_qbraAket_k=0_3}) is reduced to 
$|\psi(t+dt) \rangle = e^{-\frac{i}{\hbar} \hat{H} dt}|\psi(t) \rangle$. 
Thus, starting from eq.(\ref{time_dev_qbraAket}), we have found that the Hamiltonian $\hat{H}$ has the same form as that in the RAT. 
In addition, we have derived the Schr\"{o}dinger equation. 
Such a derivation of the Schr\"{o}dinger equation is well known in the RAT~\cite{FPIbook}.

\subsection{The derivation of the Lagrangian and momentum relation}\label{derivation_lagrangian}

Following ref.~\cite{Nagao:2011is}, 
we derive the Lagrangian and momentum relation. 
We analyze the transition amplitude 
from an initial state 
$| i \rangle$ at time $t_i$ to a final state $| f \rangle$ at time $t_f$, 
which is written as 
\begin{eqnarray}
&& \langle f |   e^{ - \frac{i}{\hbar} \hat{H} (t_f - t_i ) }  | i \rangle \nonumber \\
&=& \int_C dq_1 \cdots dq_{N}~  
\langle f | q_N \rangle_{new} 
~{}_m \langle_{new}~ q_N |  e^{ - \frac{i}{\hbar} \hat{H} \Delta t } | q_{N-1} \rangle_{new}~ 
~{}_m \langle_{new}~ q_{N-1} | \cdots | q_2 \rangle_{new}~ \nonumber \\
&&
\times
~{}_m \langle_{new}~ q_2 | e^{ - \frac{i}{\hbar} \hat{H} \Delta t } | q_1 \rangle_{new}~ 
~{}_m \langle_{new}~ q_1 | i \rangle ,  \label{braqbqaket}
\end{eqnarray}
where we have divided the time interval $t_f -t_i$ into $N-1$ pieces whose interval is 
$\Delta t = \frac{ t_f - t_i } {N-1}$, and defined $\dot{q}_j$ by 
$\dot{q_j} \equiv \frac{ q_{j+1} - q_j }{ \Delta t}$. 
Then, since ${}_m \langle_{new}~ q_{j+1} | e^{ - \frac{i}{\hbar} \hat{H} \Delta t } | q_j \rangle_{new}$ is rewritten as 
\begin{eqnarray}
{}_m \langle_{new}~ q_{j+1} | e^{- \frac{i}{\hbar}  H(\hat{p}_{new} , \hat{q}_{new}) \Delta t  } 
| q_{j} \rangle_{new}   
&=&  \int_C dp_j 
e^{- \frac{i}{\hbar}   H(p_j, q_j)  \Delta t  } ~{}_m \langle_{new}~ q_{j+1} | p_j \rangle_{new} 
~{}_m \langle_{new}~ p_j | q_{j} \rangle_{new}~ \nonumber \\
&=& 
\int_C \frac{dp_j}{2\pi\hbar} 
\exp\left[  \frac{i}{\hbar} \Delta t L(p_j, q_j, \dot{q}_{j}) \right] , \label{braqj+1qjket}
\end{eqnarray}
where $L(p_j, q_j, \dot{q}_{j})$ is given by 
\begin{eqnarray}
L(p_j, q_j, \dot{q}_{j}) &=&p_j \dot{q}_{j} - H(p_j, q_j) \nonumber \\
&=& 
-\frac{1}{2m}  \left( p_j - m \dot{q}_j \right)^2 
+ \frac{ 1}{2} m \dot{q_j}^2-V(q_j) , \label{L_saddlepoint_p}
\end{eqnarray}
the transition amplitude 
$\langle f |   e^{ - \frac{i}{\hbar} \hat{H} (t_f - t_i ) }  | i \rangle$ is 
estimated as 
\begin{eqnarray}
&&\langle f |   e^{ - \frac{i}{\hbar} \hat{H} (t_f - t_i ) }  | i \rangle \nonumber \\
&=&
\int_C \frac{dp_1}{2\pi\hbar} \cdots \frac{dp_{N-1}}{2\pi\hbar} dq_1 \cdots dq_{N}~ 
\langle f | q_N \rangle_{new}~  ~{}_m \langle_{new}~ q_1 | i \rangle
\exp\left[  \frac{i}{\hbar} \sum_{j=1}^{N-1} \Delta t  L(p_j, q_j, \dot{q}_j) \right] \nonumber \\
&=& \int_C  {\cal D} p {\cal D} q~ \psi_f(q_f)^{*_{q_f}} \psi_i(q_i)
\exp\left[  \frac{i}{\hbar} \int_{t_i}^{t_f} dt L(p, q, \dot{q})  \right] ,
\end{eqnarray}
where in the second equality we have introduced $q_i=q_1$ and $q_f=q_N$. 
We perform the following Gaussian integral 
around the saddle point $p_j=m \dot{q_j}$, 
\begin{eqnarray}
\int_C \frac{dp_j}{2\pi\hbar}  \exp\left[ \frac{i}{\hbar} \Delta t L(p_j, q_j, \dot{q}_j) \right] 
&=& 
\int_C \frac{dp_j}{2\pi\hbar} 
\exp\left[ \frac{i}{\hbar} \Delta t 
\left\{ -\frac{1}{2m}  \left( p_j - m \dot{q}_j \right)^2 
+ \frac{ 1}{2} m \dot{q_j}^2-V(q_j) \right\} \right] \nonumber \\
&=&\sqrt{ \frac{m}{2\pi i \hbar \Delta t} } \exp\left[  \frac{i}{\hbar} \Delta t L(\dot{q}_j , q_j)  \right] , 
\label{gauss_integral_p_L}
\end{eqnarray}
where $L(\dot{q_j}, q_j) = \frac{1}{2} m \dot{q_j}^2 - V(q_j)$. 
Thus, we have obtained the momentum relation (\ref{def_p_delLdelqdot}) 
and the Lagrangian (\ref{lagrangian}).

\section{Properties of the future-included theory}\label{sec:prop_future-included}

\subsection{Future-included theory}

Improving the definition given in ref.\cite{Bled2006}, 
based on the complex coordinate formalism\cite{Nagao:2011za}, 
in ref.\cite{Nagao:2012mj} we introduced $| A(t) \rangle$ and $| B(t) \rangle$ by 
\begin{eqnarray}
&&\psi_A(q)={}_m \langle_{new}~ q | A(t) \rangle = \int_{\text{path}(t)=q} 
e^{\frac{i}{\hbar} S_{T_A ~\text{to} ~t}(path)} D\text{path}, \label{psi_A(q)} \\
&&\psi_B(q)^{*_q} = \langle B(t) | q \rangle_{new} 
= \int_{\text{path}(t)=q} e^{\frac{i}{\hbar} S_{t ~\text{to} ~T_B}(path)}  D\text{path} ,
\end{eqnarray}
where $\text{path}(t)=q$ means the boundary condition at the present time $t$, 
and $T_A$ and $T_B$ are taken as $-\infty$ and $\infty$ respectively. 
$| A(t) \rangle$ and $| B(t) \rangle$ are supposed to time-develop according to  
\begin{eqnarray}
&&i \hbar \frac{d}{d t} | A(t) \rangle = \hat{H} | A(t) \rangle , 
\label{schro_eq_Astate} \\
&&i \hbar \frac{d}{d t} | B(t) \rangle = \hat{H}_B | B(t) \rangle , 
\label{schro_eq_Bstate} 
\end{eqnarray}
where $\hat{H}_B= \hat{H}^\dag$.

The authors of ref.\cite{Bled2006} 
speculated that 
the following matrix element\footnote{
In the RAT the matrix element 
$\langle \hat{\cal O} \rangle^{BA}$ is called the weak value\cite{AAV} and 
has been intensively studied. 
For details of the weak value, 
see the reviews\cite{review_wv,AT,AV08,Shikano2011review} and the references therein.} 
of some operator $\hat{\cal O}$ 
\begin{equation}
\langle \hat{\cal O} \rangle^{BA} \equiv 
\frac{ \langle B(t) |  \hat{\cal O}  | A(t) \rangle }{ \langle B(t) | A(t) \rangle } 
\label{OBA}
\end{equation}
corresponds to the expectation value in the future-not-included theory, 
\begin{equation}
\langle \hat{\cal O} \rangle^{AA} 
\equiv \frac{  \langle A(t) | \hat{\cal O}  | A(t) \rangle }
{ \langle A(t) | A(t) \rangle } ,  \label{OAA}
\end{equation}
i.e. $\langle \hat{\cal O} \rangle^{BA} \simeq \langle \hat{\cal O} \rangle^{AA}$. 
In refs.\cite{Nagao:2012mj,Nagao:2012ye} 
we investigated $\langle \hat{\cal O} \rangle^{BA}$ carefully. 
Using both the complex coordinate formalism\cite{Nagao:2011za} 
and the automatic hermiticity mechanism\cite{Nagao:2010xu,Nagao:2011za}, 
i.e., a mechanism to obtain the Hermitian Hamiltonian 
after a long time development, 
we obtained a correspondence principle that 
$\langle \hat{\cal O} \rangle^{BA}$ for large $T_B-t$ and large $t-T_A$ is almost 
equivalent to $\langle \hat{\cal O} \rangle_{Q'}^{AA}$ for large $t-T_A$, 
where $Q'$ is a Hermitian operator which is used to define a proper 
inner product.\footnote{For simplicity, in this paper we are not concerned with the proper 
inner product, which is defined by making the Hamiltonian normal, 
since it does not have an essential role in this study. }

We note that $\langle \hat{\cal O} \rangle^{BA}$ is not an expectation value but a matrix 
element in the usual sense. 
But in ref.\cite{Nagao:2012mj} 
we found that if we regard it as an expectation value in the future-included theory, 
then we obtain the Heisenberg equation, Ehrenfest's theorem and a conserved probability current density. 
This result strongly suggests that we can regard $\langle \hat{\cal O} \rangle^{BA}$ 
as an expectation value in the future-included theory.

\subsection{The Heisenberg equation and Ehrenfest's theorem} 

In ref.\cite{Nagao:2012mj} 
we defined the Heisenberg operator, 
\begin{equation}
\hat{\cal O}_H^{fi}(t , t_\text{ref}) \equiv 
\exp\left( \frac{i}{\hbar} \hat{H} (t- t_\text{ref}) \right) 
\hat{\cal O}  \exp\left( -\frac{i}{\hbar} \hat{H} (t-t_\text{ref}) \right), 
\label{OTAHeisenbergop}
\end{equation}
where $\hat{H}$ is given in eq.(\ref{expHhat}) and 
$t_\text{ref}$ is some reference time 
chosen arbitrarily such that $T_A \le t_\text{ref} \le T_B$. 
This Heisenberg operator, which appears in the numerator of 
$\langle \hat{\cal O} \rangle^{BA}$ as 
$\langle B(t) |  \hat{\cal O}  | A(t) \rangle =
\langle B(t_\text{ref}) | \hat{\cal O}_H^{fi}(t , t_\text{ref}) | A(t_\text{ref}) \rangle$, 
obeys the Heisenberg equation 
\begin{equation}
\frac{d}{dt}  \hat{\cal O}_H^{fi}(t , t_\text{ref})
= \frac{i}{\hbar} [ \hat{H} , \hat{\cal O}_H^{fi}(t , t_\text{ref}) ]. \label{Hei_eq_fi}
\end{equation}
In addition, 
since $\langle \hat{\cal O} \rangle^{BA}$ obeys 
\begin{eqnarray}
\frac{d}{dt} \langle \hat{\cal O} \rangle^{BA} 
&=& \langle  \frac{i}{\hbar} [ \hat{H} , \hat{\cal O} ]   \rangle^{BA} , 
\label{time_dev_OBA}
\end{eqnarray} 
we obtain 
\begin{eqnarray}
&&\frac{d}{dt} \langle \hat{q}_{new} \rangle^{BA} 
= \frac{1}{m} \langle \hat{p}_{new} \rangle^{BA} ,  \label{dqcdt}  \\
&&\frac{d}{dt} \langle \hat{p}_{new} \rangle^{BA} 
= - \langle V'(\hat{q}_{new}) \rangle^{BA} ,    \label{dpcdt} 
\end{eqnarray}
and Ehrenfest's theorem, 
$m\frac{d^2}{dt^2} \langle \hat{q}_{new} \rangle^{BA} 
= - \langle V'(\hat{q}_{new}) \rangle^{BA}$. 
Thus, $\langle \hat{\cal O} \rangle^{BA}$ provides 
the time development of the saddle point for $\exp(\frac{i}{\hbar} S)$. 
Since eq.(\ref{dqcdt}) is consistent with eq.(\ref{def_p_delLdelqdot}), 
eq.(\ref{def_p_delLdelqdot}) 
is confirmed to be the momentum relation in the future-included theory.

\section{Properties of the future-not-included theory}\label{sec:future-not-included}

\subsection{The Heisenberg and Schr\"{o}dinger equations} 

Following refs.\cite{Nagao:2010xu,Nagao:2011za}, 
we explain the time development of 
$\langle \hat{\cal O} \rangle^{AA}$ given in eq.(\ref{OAA}). 
Introducing a normalized state $|A(t) \rangle_{N}$ by 
\begin{equation}
|A(t) \rangle_{N} \equiv \frac{1}{\sqrt{ \langle {A}(t) | ~{A}(t) \rangle} } | {A}(t) \rangle , 
\end{equation} 
we express $\langle \hat{\cal O} \rangle^{AA}$ as 
\begin{eqnarray}
\langle \hat{\cal O} \rangle^{AA}  
&=& {}_{N} \langle A(t) | \hat{\cal O} | A(t) \rangle_{N} \nonumber \\
&=&  {}_{N} \langle A(t_0) | \hat{\cal O}_{H}^{fni}(t, t_0) | A(t_0) \rangle_{N}, 
\end{eqnarray} 
where we have introduced the Heisenberg operator $\hat{\cal O}_{H}^{fni}(t, t_0)$ by 
\begin{equation}
\hat{\cal O}_{H}^{fni}(t, t_0) \equiv \frac{ \langle A(t_0) | A(t_0) \rangle }{ \langle A(t) | A(t) \rangle } 
e^{ \frac{i}{\hbar} \hat{H}^{\dag} (t-t_0) } \hat{\cal O} e^{ -\frac{i}{\hbar} 
\hat{H}(t-t_0)}.  
\end{equation}
This operator $\hat{\cal O}_{H}^{fni}(t, t_0)$ obeys the slightly modified Heisenberg equation, 
\begin{eqnarray}
i\hbar  \frac{d}{d t} \hat{\cal O}_{H}^{fni}(t, t_0) 
&=& \hat{\cal O}_{H}^{fni}(t, t_0) \hat{H} - \hat{H}^{\dag} \hat{\cal O}_{H}^{fni}(t, t_0) 
-2 {}_{N} \langle A(t) | \hat{H}_{a} | A(t) \rangle_{N} 
\hat{\cal O}_{H}^{fni}(t, t_0) \nonumber \\
&=& [ \hat{\cal O}_{H}^{fni}(t, t_0) , \hat{H}_{h} ] 
+ \left\{  \hat{\cal O}_{H}^{fni}(t, t_0) , \hat{H}_{a} 
- {}_{N} \langle A(t) | \hat{H}_{a} | A(t) \rangle_{N} \right\} , \label{Hei_eq_fni}
\end{eqnarray}
where $\hat{H}_{h}$ and $\hat{H}_{a}$ are the Hermitian and anti-Hermitian parts of 
$\hat{H}$ respectively. 
We note that eq.(\ref{Hei_eq_fni}) is more complicated than 
the Heisenberg equation in the future-included theory, eq.(\ref{Hei_eq_fi}). 
In addition, $| A(t) \rangle_{N}$ obeys 
the slightly modified Schr\"{o}dinger equation,
\begin{eqnarray}
i\hbar  \frac{d}{d t} | A(t) \rangle_{N} 
&=& \hat{H} | A(t) \rangle_{N} 
-{}_{ N} \langle A(t) | \hat{H}_{a} | A(t) \rangle_{N} | A(t) \rangle_{N} \nonumber \\
&=& \hat{H}_{h} | A(t) \rangle_{N} 
+ \left( \hat{H}_{a} -{}_{N} \langle A(t) | \hat{H}_{a} | A(t) \rangle_{N} \right) 
| A(t) \rangle_{N} . \label{sch_fni}
\end{eqnarray} 
%

\subsection{Classical limit of the future-not-included theory}

As we pointed out in refs.\cite{Nagao:2010xu,Nagao:2011za}, 
eqs.(\ref{Hei_eq_fni})(\ref{sch_fni}) suggest that 
the effect of the anti-Hermitian part of the non-Hermitian Hamiltonian $\hat{H}$ 
disappears in the classical limit, 
though the theory is defined with $\hat{H}$ at the quantum level. 
To see this in terms of the expectation value 
$\langle \hat{\cal O} \rangle^{A A}$, utilizing eq.(\ref{sch_fni})
we give the following expression, 
\begin{eqnarray}
i \hbar \frac{d}{d t} \langle \hat{\cal O}  \rangle^{A A} 
&=&
\langle [\hat{\cal O}, \hat{H}_h] \rangle^{A A} + \langle F(\hat{\cal O}, \hat{H}_a) 
\rangle^{A A} , \nonumber \\
&\simeq&
\langle [\hat{\cal O}, \hat{H}_h] \rangle^{A(t) A(t)} ,  \label{ihbardeldeltOAA_t}
\end{eqnarray}
where $F(\hat{\cal O}, \hat{H}_a)(t)$, a quantum fluctuation term given by 
\begin{eqnarray}
F(\hat{\cal O}, \hat{H}_a)(t) 
&=& 
\left\{ \hat{\cal O} , \hat{H}_a  - \langle \hat{H}_a \rangle^{A A} \right\}  \nonumber \\
&=& 
\left\{ \hat{\cal O} -\langle\hat{\cal O} \rangle^{A A} , \hat{H}_a  \right\},  
\label{fluc_term_O_Ha}
\end{eqnarray}
disappears in the classical limit, so we have used the relation 
$\langle F(\hat{\cal O}, \hat{H}_a) \rangle^{AA} \simeq 0$.

Substituting $\hat{q}_{new}$ and $\hat{p}_{new}$ 
for $\hat{\cal O}$ in eq.(\ref{ihbardeldeltOAA_t}), we obtain 
\begin{eqnarray} 
\frac{d}{d t} \langle \hat{q}_{new}  \rangle^{A A} 
&\simeq& \frac{1}{ i \hbar   }
\langle [\hat{q}_{new} , \hat{H}_h] \rangle^{A A}   \nonumber \\
&\simeq& \frac{1}{m_{\text{eff}}} 
\langle \hat{p}_{new} \rangle^{A A}  ,  \label{ihbardeldeltqhatAA_t} \\ 
\frac{d}{dt} \langle \hat{p}_{new}  \rangle^{A A} 
&\simeq&
\langle [\hat{p}_{new} , \hat{H}_h] \rangle^{A A}  \nonumber \\
&\simeq&
- \langle V_R'( \hat{q}_{new} ) \rangle^{A A}   ,   \label{ihbardeldeltphatAA_t}
\end{eqnarray}
where, in the last line of each relation, $m_{\text{eff}}$ and $V_R$ are given by 
\begin{equation}
m_{\text{eff}} \equiv m_R + \frac{m_I^2}{m_R}  \label{m_eff} 
\end{equation}
and eq.(\ref{V_R}), and we have taken into account Theorem 1  
given in subsection~\ref{prop_qnewpnew} and used 
the approximation that $\hat{q}_{new} \simeq \hat{q}$ and 
$\hat{p}_{new} \simeq \hat{p}$. 
Since eq.(\ref{ihbardeldeltqhatAA_t}) suggests the following momentum relation, 
\begin{equation}
p= m_{\text{eff}} \dot{q}, \label{momentum_fni}
\end{equation}
we claim that this is the momentum relation in the future-not-included theory. 
Eq.(\ref{momentum_fni}) is different from eq.(\ref{def_p_delLdelqdot}), 
which is confirmed to be the momentum relation in the future-included theory. 
But, in the future-not-included theory, 
where both $\langle \hat{q}_{new} \rangle^{A A}$ and 
$\langle \hat{p}_{new} \rangle^{A A}$ 
are real for $\hat{q}_{new}$ and $\hat{p}_{new}$ replaced with Hermitian 
$\hat{q}$ and $\hat{p}$ respectively, 
eq.(\ref{def_p_delLdelqdot}) is inconsistent 
because $m$ is complex. 
On the other hand, 
we do not encounter such a contradiction for eq.(\ref{momentum_fni}) 
in the future-not-included theory, because $m_{\text{eff}}$ is real. 
Therefore, we conclude that the momentum relations 
in the future-included and future-not-included theories 
are given by eqs.(\ref{def_p_delLdelqdot})(\ref{momentum_fni}), respectively. 
Then, one may question why the method of ref.\cite{Nagao:2011is} 
for deriving eq.(\ref{def_p_delLdelqdot}), which was explained 
in section~\ref{sec:review_xi-state_argument}, 
does not work in the future-not-included theory. 
Later, in section 6, we will come back to this point 
and explain that the method works 
even in the future-not-included theory, and provides eq.(\ref{momentum_fni}), 
if it is properly applied to the future-not-included theory.

Combining eq.(\ref{ihbardeldeltqhatAA_t}) with eq.(\ref{ihbardeldeltphatAA_t}), 
we obtain Ehrenfest's theorem, 
\begin{equation}
m_{\text{eff}} \frac{d^2}{d t^2} \langle \hat{q}_{new}  \rangle^{A A}  
\simeq - \langle V_R'( \hat{q}_{new} ) \rangle^{A A}, 
\end{equation}
which suggests that the classical theory of the future-not-included theory 
is described not by a full action $S$, but $S_{\text{eff}}$ defined by 
\begin{eqnarray}
S_{\text{eff}}&\equiv&\int_{T_A}^t dt L_{\text{eff}} , \label{Seff} \\
L_{\text{eff}}(\dot{q}, q) 
&\equiv& \frac{1}{2} m_{\text{eff}} \dot{q}^2 - V_{R}(q) .
\end{eqnarray}
Here we note that 
$L_{\text{eff}}$ is different from $L_R$ given in eq.(\ref{L_R}). 
Thus, we claim that the classical theory of the future-not-included theory 
is described by $\delta S_{\text{eff}} = 0$. 
Then the momentum relation given in eq.(\ref{momentum_fni}) 
is rewritten as $p=\frac{\partial L_{\text{eff}}}{\partial \dot{q}}$. 
This is quite in contrast to the classical theory of the future-included theory, 
which would be described by $\delta S = 0$, where $S=\int_{T_A}^{T_B} dt L$, 
and the momentum relation given by eq.(\ref{def_p_delLdelqdot}). 
In addition, the classical Hamiltonian in the future-not-included theory 
is given by 
\begin{eqnarray}
H_R \equiv \text{Re}_q H 
= \frac{1}{2m_{\text{eff}}}  p^2 + V_R(q), \label{effective_classicalH0}  
\end{eqnarray}
where $H_R$ is the $q$-real part of the classical Hamiltonian 
$H \equiv \frac{1}{2m} p^2 + V(q)$,  
which is given by replacing $\hat{q}_{new}$ and $\hat{p}_{new}$ 
with $q$ and $p$ respectively in $\hat{H}$. 
In refs.\cite{Nagao:2010xu,Nagao:2011za}  
introducing a proper inner product so that the eigenstates of 
$\hat{H}$ are orthogonal to each other and considering a long time development, 
we obtained a Hermitian Hamiltonian. 
But now without using the automatic hermiticity mechanism 
we have obtained a real Hamiltonian in the classical limit. 
This is an intriguing property of the future-not-included theory,  
though restricted to the classical limit. 
We make a comparison between the future-included and future-not-included theories 
in Table~\ref{tab:comparison_fi_fni}.

\begin{table}
\caption[AAA]{Comparison between the future-included and future-not-included theories}
\label{tab:comparison_fi_fni}
\begin{center}
\begin{tabular}{|p{3.5cm}|p{4cm}|p{7cm}|}
\hline
    & future-included theory & future-not-included theory \\
\hline
action  &  $S=\int_{T_A}^{T_B}dt L$   & $S=\int_{T_A}^{t} dt L$ \\  
\hline
``expectation value" & $\langle \hat{\cal O} \rangle^{BA} = 
\frac{ \langle B(t) |  \hat{\cal O}  | A(t) \rangle }{ \langle B(t) | A(t) \rangle }$ & $\langle \hat{\cal O} \rangle^{AA} 
= \frac{  \langle A(t) | \hat{\cal O}  | A(t) \rangle }
{ \langle A(t) | A(t) \rangle }$  \\   
\hline
time development 
& $i \hbar \frac{d}{d t} \langle \hat{\cal O} \rangle^{BA}$  
& $i \hbar \frac{d}{d t} \langle \hat{\cal O}  \rangle^{AA}$ \\
& $= \langle  [ \hat{\cal O} , \hat{H} ]\rangle^{BA}$
& 
$= \langle [\hat{\cal O}, \hat{H}_h] \rangle^{AA} 
+ \langle \left\{ \hat{\cal O} -\langle\hat{\cal O} \rangle^{A A} , \hat{H}_a  \right\} \rangle^{A A}$  
\\ 
&
&
$\simeq \langle [\hat{\cal O}, \hat{H}_h] \rangle^{AA} $ \\
\hline
classical theory & $\delta S=0$ &  $\delta S_{\text{eff}}=0$,  $S_{\text{eff}}=\int_{T_A}^{t} dt L_{\text{eff}}$\\  
\hline
momentum relation  &  $p= m \dot{q}$   & $p= m_{\text{eff}} \dot{q}$ \\    
\hline
\end{tabular}
\end{center}
\end{table}
%
%

\subsection{Another method for seeing the time development of $\langle \hat{\cal O} \rangle^{AA}$ by re-choosing the $B$ state}\label{subsec:rechoosing}

The quantity $\langle \hat{\cal O} \rangle^{BA}$ in the future-included theory 
behaves as an expectation value, despite looking like a matrix element, 
and it time-develops according to the very simple 
expression of eq.(\ref{time_dev_OBA}). 
On the other hand, the expectation value $\langle \hat{\cal O} \rangle^{AA}$ 
in the future-not-included theory time-develops in a more complicated way 
at the quantum level with the additional term 
$\langle \left\{ \hat{\cal O}, \hat{H}_a-\langle \hat{H}_a \rangle \right\} \rangle^{AA}$, 
as seen in eq.(\ref{ihbardeldeltOAA_t}). 
Hence, we are motivated to 
study whether we can simplify the description of the time development of 
$\langle \hat{\cal O} \rangle^{AA}$ 
by rewriting it formally in the expression of the future-included theory 
and utilizing the simple time development of the future-included theory. 
Even if we cannot make it simpler, 
it would be interesting to reproduce and understand 
the time development of the future-not-included theory  
from a different point of view via the future-included theory. 
At the least, this would become a consistency check of the theory, 
and we could claim that the future-included theory can be used as a 
mathematical tool to compute the time development of $\langle \hat{\cal O} \rangle^{AA}$. 
Therefore, in this subsection, 
we try to describe the time development of the expectation value of the future-not-included theory $\langle \hat{\cal O} \rangle^{AA}$ 
by making use of the future-included theory.

We begin by putting the condition 
\begin{equation}
\langle q |  B(t) \rangle = \langle q | A(t) \rangle  \label{cond_qBqA_t}
\end{equation}
on the $B$ state at some time $t$.\footnote{We cannot simply use eq.(\ref{cond_qBqA_t}) 
except for at one value of $t$, because 
the states $| A(t) \rangle$ and $| B(t) \rangle$ time-develop differently: 
according to eqs.(\ref{schro_eq_Astate})(\ref{schro_eq_Bstate}), respectively. 
} 
We call this ``re-choosing" the $B$ state. 
Expressing the $B$ state re-chosen at $t$ as $| B_{t}(t') \rangle$, 
where $t'$ is a formal time to allow the time-development as a $B$ state, 
we have the following relation for the time $t$: 
\begin{equation}
| B_{t}(t) \rangle = | A(t) \rangle. \label{Bttket_Atket}
\end{equation}
Then eq.(\ref{OAA}) is rewritten as 
\begin{eqnarray}
\langle \hat{\cal O} \rangle^{AA} 
&=& \frac{\langle B_t(t) | \hat{\cal O} | A(t) \rangle}{ \langle B_t(t) | A(t) \rangle } 
\equiv 
\langle \hat{\cal O} \rangle^{B_t A}    
\end{eqnarray}
for each $t$. 
In a realistic future-included theory it would be a very strange accident 
to have the relation of eq.(\ref{cond_qBqA_t}) even at one time. 
Hence, the re-choosing cannot be taken seriously. 
We just look for some formal rule to use the future-included theory 
as long as possible but to obtain the future-not-included theory as our result.

The re-chosen $B$ state $| B_{t}(t') \rangle$ obeys 
\begin{eqnarray}
&& i \hbar \frac{d}{dt}  | B_t (t) \rangle = H | B_t (t) \rangle , \label{AstateBttstate} \\
&& i \hbar\frac{\partial}{\partial t'}  | B_{t} (t') \rangle = H^\dag | B_{t} (t') \rangle , \label{BstateBttstate}
\end{eqnarray}
which come from eqs.(\ref{schro_eq_Astate})(\ref{schro_eq_Bstate}) respectively. 
Using eqs.(\ref{Bttket_Atket})(\ref{BstateBttstate}), we can calculate 
the time derivative of $| A(t) \rangle$ as 
\begin{eqnarray}
\frac{d}{dt}  | A(t) \rangle 
&=& \left( \frac{\partial}{\partial t}  | B_{t} (t') \rangle \right) |_{t'=t} 
- \frac{i}{\hbar} H^\dag | B_{t} (t) \rangle . 
\end{eqnarray}
Since eq.(\ref{BstateBttstate}) provides the expression 
\begin{eqnarray}
| B_t(t') \rangle 
&=& e^{- \frac{i}{\hbar} H^\dag(t' - t'')} | B_t(t'') \rangle
= e^{- \frac{i}{\hbar} H^\dag(t' - t)} | B_t(t) \rangle, 
\end{eqnarray}
we obtain 
\begin{eqnarray}
\frac{\partial}{\partial t} | B_t(t') \rangle 
&=& \frac{i}{\hbar} \left( H^\dag | B_t(t') \rangle 
-  e^{- \frac{i}{\hbar} H^\dag (t' - t)} H | B_t(t) \rangle \right) . 
\end{eqnarray}
For $t'=t$ this is expressed as 
\begin{eqnarray}
i \hbar \left( \frac{\partial}{\partial t} | B_t(t') \rangle \right) |_{t'=t}
&=&( H - H^\dag ) | B_t(t') \rangle|_{t'=t} \nonumber \\
&=&2H_{a} e^{-\frac{i}{\hbar} H^\dag (t-t'') } | B_t(t'') \rangle, \label{ihbarddtBtt}
\end{eqnarray}
where the left-hand side is rewritten as 
\begin{eqnarray}
i \hbar \left\{ \frac{\partial}{\partial t} \left( e^{-i H^\dag (t'-t'') } | B_t(t'') \rangle \right) \right\} |_{t'=t}
&=&  e^{-\frac{i}{\hbar} H^\dag (t-t'') } i \hbar \frac{\partial}{\partial t} | B_t(t'') \rangle .
\end{eqnarray}
Therefore, we obtain 
\begin{eqnarray}
i \hbar \frac{\partial}{\partial t} | B_t(t'') \rangle 
&=& {\cal U}_{t'', t}^{-1}  2H_{a} {\cal U}_{t'', t} | B_t(t'') \rangle, 
\label{time_t_devlop_Bt(t'')}
\end{eqnarray}
where we have introduced 
\begin{equation}
{\cal U}_{t'', t} = e^{-\frac{i}{\hbar} H^\dag (t-t'') } .
\end{equation}

Next, we calculate the time derivative of $\langle \hat{\cal O}  \rangle^{A A}$,  
\begin{eqnarray}
\frac{\partial}{\partial t} \langle \hat{\cal O}  \rangle^{A(t) A(t)}
&=&
\left\{ \frac{\partial}{\partial t} \langle \hat{\cal O}  \rangle^{B_{t}(t') A(t')}  \right\}|_{t'=t}
+
\left\{ \frac{\partial}{\partial t'} \langle \hat{\cal O}  \rangle^{B_{t}(t') A(t')}  \right\}|_{t'=t} , 
\label{deldelt'OBt'At_t'=t}
\end{eqnarray}
where 
$\langle \hat{\cal O}  \rangle^{B_t(t') A(t')} 
= \frac{\langle B_t(t') | \hat{\cal O} | A(t') \rangle}{ \langle B_t(t') |A(t') \rangle}$ 
is formally a good classical solution in the future-included theory 
for each $t'$ as long as the equation of motion is considered. 
Indeed, the second term of eq.(\ref{deldelt'OBt'At_t'=t}) is expressed as 
\begin{equation}
\left\{ \frac{\partial}{\partial t'} \langle \hat{\cal O}  \rangle^{B_t(t') A(t')} 
\right\}|_{t'=t}
=
\frac{i}{\hbar} \langle [H, \hat{\cal O}] \rangle^{B_t(t') A(t')}|_{t'=t}. 
\label{deldelt'O} 
\end{equation}
On the other hand, the first term of eq.(\ref{deldelt'OBt'At_t'=t}) 
does not become a simple expression. We can rewrite this by utilizing 
eq.(\ref{time_t_devlop_Bt(t'')}) as follows: 
\begin{eqnarray}
\left\{ \frac{\partial}{\partial t} \langle \hat{\cal O}  \rangle^{B_{t}(t') A(t')} 
\right\}|_{t'=t}
&=&
\left\{
-\frac{2i}{\hbar} \frac{1}{ \langle B_{t}(t') | A(t') \rangle } 
\langle B_{t}(t') | U_{t',t}^\dag H_a (U_{t',t}^{-1})^\dag \hat{\cal O} | A(t') \rangle 
\right. \nonumber \\
&&
+ \left.
\frac{2i}{\hbar} \frac{ \langle B_{t}(t') | \hat{\cal O} | A(t') \rangle }{ ( \langle B_{t}(t') | A(t') \rangle )^2 } 
\langle B_{t}(t') | U_{t',t}^\dag H_a (U_{t',t}^{-1})^\dag | A(t') \rangle \right\}|_{t'=t} \nonumber \\
&=&
\frac{2i}{\hbar}
\left\{
\langle \hat{\cal O}  \rangle^{A(t) A(t)}  
\langle H_a  \rangle^{A(t) A(t)}  
-
\langle H_a \hat{\cal O}  \rangle^{A(t) A(t)}  
\right\} , \nonumber \\
&=&
\frac{1}{i \hbar} 
\left[
- \langle [\hat{\cal O}, H_a] \rangle^{A(t) A(t)}  
+ \langle \left\{ \hat{\cal O} -\langle\hat{\cal O} \rangle^{A(t) A(t)} , H_a  \right\} 
\rangle^{A(t) A(t)}  
\right] .  \nonumber \\ \label{deldeltO} 
\end{eqnarray}
Substituting eqs.(\ref{deldelt'O})(\ref{deldeltO}) 
for eq.(\ref{deldelt'OBt'At_t'=t}), we obtain eq.(\ref{ihbardeldeltOAA_t}). 
Thus, we have shown that we can derive 
the time development of $\langle \hat{\cal O} \rangle^{AA}$, 
the expectation value in the future-not-included theory, 
by making use of the future-included theory. 
In particular, we have explicitly seen that 
it is the first term of eq.(\ref{deldelt'OBt'At_t'=t}) 
that provides the anti-commutator term, 
which disappears in the classical limit, besides the commutator 
$\langle [\hat{\cal O}, H_a] \rangle^{A(t) A(t)}$. 
As a result, this method is not so simple, but it is interesting 
in the sense that this provides another way to understand 
the time development of the future-not-included theory. Indeed, we have seen that 
the time development of $\langle \hat{\cal O} \rangle^{AA}$ 
is expressed as the simple time development of $\langle \hat{\cal O} \rangle^{BA}$ 
and a slightly complicated correction due to the formal re-choosing of the $B$ state.

\section{Reconsideration of the method for deriving the momentum relation via the Feynman path integral in the future-not-included theory}\label{sec:reconsideration}

In the foregoing sections we have seen that 
the momentum relation of eq.(\ref{def_p_delLdelqdot}) derived via FPI 
in ref.\cite{Nagao:2011is} is valid in the future-included theory, 
because it is consistent with eq.(\ref{dqcdt}), which was derived 
by looking at the time development of $\langle \hat{q}_{new} \rangle^{BA}$ 
in the future-included theory. 
In eq.(\ref{momentum_fni}) we obtained another momentum relation 
in the future-not-included theory 
by analyzing the time development of $\langle \hat{q}_{new} \rangle^{AA}$. 
Now, one might question why the method of ref.\cite{Nagao:2011is} 
for deriving the momentum relation via FPI, which was 
reviewed in section~\ref{sec:review_xi-state_argument}, 
is not valid in the future-not-included theory. 
The reason is as follows: 
In the method of ref.\cite{Nagao:2011is}, 
we analyzed the time development of 
a $\xi$-parametrized state in a transition amplitude 
from the initial time $t_i$ to the final time $t_f$, where the present time 
$t$ is supposed to be between $t_i$ and $t_f$. 
Such a transition amplitude is similar to that  
in the future-included theory, which is written as 
\begin{equation}
\langle B(t)| A(t) \rangle 
= 
\langle B(T_B)|  
\exp\left(-\frac{i}{\hbar} \hat{H}(T_B-T_A) \right) | A(T_A) \rangle , 
\end{equation}
where the present time $t$ is between $T_A$ and $T_B$.  
On the other hand, 
in the future-not-included theory the transition amplitude is given by 
\begin{equation}
\langle A(t)| A(t) \rangle 
= 
\langle A(T_A)| \exp\left( \frac{i}{\hbar} \hat{H}^\dag(t-T_A) \right) 
\exp\left(-\frac{i}{\hbar} \hat{H}(t-T_A) \right) | A(T_A) \rangle , \label{AbraAket}
\end{equation}
so we have to consider a path starting from 
the initial time $T_A$ to the present time $t$, and 
also that going backward from $t$ to $T_A$. 
In this section we discuss how to apply the method of ref.\cite{Nagao:2011is}  
for deriving the momentum relation via the Feynman path integral, 
which was reviewed in section~\ref{sec:review_xi-state_argument},  
to the future-not-included theory.

\subsection{Formal Lagrangian in the future-not-included theory}

To apply the method of ref.\cite{Nagao:2011is} to the future-not-included theory, 
we formally rewrite the transition amplitude 
$\langle A(t)| A(t) \rangle$ into another expression 
similar to $\langle B(t)| A(t) \rangle$, 
and introduce a formal Lagrangian $L_{\text{formal}}$. 
We argue that using this formal Lagrangian $L_{\text{formal}}$ 
in place of the original Lagrangian $L$ in 
the method of ref.\cite{Nagao:2011is} 
we obtain the momentum relation in the future-not-included theory, 
eq.(\ref{momentum_fni}).

In the future-not-included theory, we can rewrite eq.(\ref{AbraAket}) as 
the following path integral 
\begin{equation}
I
\equiv \int_{C} {\cal D} q \int_{C'} {\cal D} q' 
e^{ - \frac{i}{\hbar} S_{T_A ~\text{to} ~t}(q)^{*_{q}} }  
e^{ \frac{i}{\hbar} S_{T_A ~\text{to} ~t} (q')} \psi_A(q_{T_A} , T_A)^{*_{q_{T_A}}}  \psi_A(q'_{T_A} ,T_A) . 
\end{equation}
At an intermediate time $t'$ such that $T_A < t' < t$, 
we would be allowed to use a kind of future-included formulation, 
because it looks as if there is a future for $t'$. 
But for the present time $t$ there is no future but only the past, 
so we have to be careful about quantities at the time $t$, especially $\dot{q}$, etc. 
Therefore, we transform $I$ 
into an expression like a transition amplitude from the time $T_A$ to $T_B$ 
by inverting the time direction of the transition amplitude from $T_A$ to $t$ 
so that $t$ becomes an intermediate time. 
For this purpose we express $S_{T_A ~\text{to} ~t}(q)^{*_{q}}$ as  
\begin{eqnarray}
S_{T_A ~\text{to} ~t} (q)^{*_{q}} 
&=& \int_{T_A}^t dt' L( q(t'), \dot{q}(t') )^{*_{q}} \nonumber \\
&=&  \int_t^{-T_A + 2t} dt'' 
L( q_{\text{formal}}( t'', t) , -\partial_{t''} q_{\text{formal}}( t'', t) )^{*_{q_{\text{formal}}}} , 
\label{Sstar}
\end{eqnarray}
where in the second equality we have changed the variable by 
\begin{equation}
t''=-t' + 2t , \label{time_reflection} 
\end{equation}
and introduced the formal coordinate $q_{\text{formal}}$ by 
\begin{equation}
q_{\text{formal}}( t'', t) \equiv q( -t'' + 2t) = q(t') ,   
\end{equation}
which has the time dependence of not only $t$ but also $t''$ and suggests 
\begin{eqnarray}
q(t) = q_{\text{formal}}(t, t) .  \label{qqformal} 
\end{eqnarray} 
Then $I$ is written as 
\begin{eqnarray}
I
&=&\int_{C'} {\cal D}q' \int_{C''} {\cal D}q_{\text{formal}} 
\exp\left[ \frac{i}{\hbar} \int_{T_A}^t dt' L( q'(t') , \dot{q'} (t') )  \right]
\nonumber \\
&&
\times 
\exp\left[-\frac{i}{\hbar} \int_t^{T_B} dt'' 
L( q_{\text{formal}}( t'', t), -\partial_{t''} q_{\text{formal}}( t'', t) )^{*_{q_{\text{formal}}}}  \right] 
J \psi_A(q'_{T_A}, T_A) , \label{I2}
\end{eqnarray}
where 
$C''$ is a contour of $q_{\text{formal}}( t'', t)$, 
which is obtained by a reflection of $C$ at $t$ in the time direction, 
and $J$ is given by 
\begin{eqnarray}
J 
&=&\int_{C'''} {\cal D}q'_{\text{formal}}  
\exp\left[-\frac{i}{\hbar} \int_{T_B}^{-T_A + 2t} dt'' 
L( q'_{\text{formal}}( t'', t) , -\partial_{t''} q'_{\text{formal}}( t'', t) )^{*_{q'_{\text{formal}}}}  
\right] 
\nonumber \\
&& \times
\psi_A({q'_{\text{formal}}}(-T_A + 2t, t), T_A )^{*_{q'_{\text{formal}}}} \nonumber \\
&=& 
\left\{ 
\int_{C'''} d{q'_{\text{formal}}}(-T_A+2t,t) 
~{}_m\langle {q'_{\text{formal}}}(-T_A+2t, t) 
| e^{-\frac{i}{\hbar} \hat{H}( -T_A  + 2t - T_B) } 
| {q'_{\text{formal}}}(T_B, t) \rangle 
\right. 
\nonumber \\
&& 
\left.
\times
~{}_m\langle {q'_{\text{formal}}}(-T_A+2t, t) | A(T_A) \rangle 
\right\}^{*_{q'_{\text{formal}}}}. 
\end{eqnarray} 
Using the relation 
\begin{eqnarray}
&&{}_m\langle {q'_{\text{formal}}}(-T_A+2t, t)  | 
e^{-\frac{i}{\hbar} \hat{H} ( -T_A  + 2t - T_B) } 
| {q'_{\text{formal}}}(T_B, t) \rangle \nonumber \\
&=&
~{}_m\langle {q'_{\text{formal}}}(T_B,t)    
| e^{-\frac{i}{\hbar} \hat{H} ( -T_A  + 2t - T_B) }  
|  {q'_{\text{formal}}}(-T_A+2t, t)  \rangle ,   
\end{eqnarray}
we obtain a simple expression for $J$,  
\begin{eqnarray}
J
&=& \langle A(2t - T_B) | {q'_{\text{formal}}}(T_B, t) \rangle \nonumber \\
&=&\psi_{A}( {q'_{\text{formal}}}(T_B, t) , 2t - T_B )^{*_{q'_{\text{formal}}}}. 
\end{eqnarray}
We note that the time $2t-T_B$ 
is not so far from $T_A$ because we suppose $T_B \simeq -T_A \simeq \infty$. 
Expressing $q'(t')$ for $T_A\leq t' \leq t$ as $q_{\text{formal}}(t',t)$ formally, 
we can rewrite the integral $I$ as 
\begin{eqnarray}
I &\simeq& \int {\cal D}q_{\text{formal}} \exp\left[ \frac{i}{\hbar} \int_{T_A}^{T_B} dt' 
\left\{ - \epsilon(t' -t) \right\} L_{\text{formal}}( q_{\text{formal}}(t',t), \partial_{t'} q_{\text{formal}}(t',t) , t' -t ) \right] 
\nonumber \\
&&\times \psi_{A} ( {q_{\text{formal}}}(T_B, t) , 2t - T_B )^{*_{q_{\text{formal}}}} 
\psi_A({q_{\text{formal}}}(T_A,t), T_A) , 
\label{Lformal0}
\end{eqnarray}
where $\epsilon(t)$ is a step function defined as $1$ for $t>0$ and $-1$ 
for $t<0$, and 
we have introduced the formal Lagrangian $L_{\text{formal}}$ by 
\begin{eqnarray}
&&L_{\text{formal}}( q_{\text{formal}}(t', t), \partial_{t'} q_{\text{formal}}(t', t) , t' -t ) \nonumber \\
&\equiv& 
\text{Re}_{q_{\text{formal}}} 
L( q_{\text{formal}}(t', t) , - \epsilon(t' -t) \partial_{t'} q_{\text{formal}} (t', t) )  \nonumber \\
&&
- i \epsilon(t' -t)  \text{Im}_{q_{\text{formal}}} L( q_{\text{formal}}(t',t) , - \epsilon(t' -t) \partial_{t'} q_{\text{formal}}(t',t) )  \nonumber \\
&=& \frac{1}{2} m_{\text{formal}}(t' -t) ~(\partial_{t'} q_{\text{formal}}(t',t))^2 
-V_{\text{formal}} ( q_{\text{formal}}(t',t), t' -t ) .  
\label{L_formal} 
\end{eqnarray} 
Here, 
$m_{\text{formal}}(t' -t)$ and $V_{\text{formal}}( q_{\text{formal}}(t',t), t' -t )$ 
are the formal mass and potential given by 
\begin{eqnarray}
m_{\text{formal}}(t' -t) &\equiv& m_R -  i \epsilon(t' -t) m_I , \label{m_formal} \\
V_{\text{formal}}( q_{\text{formal}}(t', t), t' -t ) &\equiv& V_R(q_{\text{formal}}(t', t))  - i \epsilon(t' -t) V_I(q_{\text{formal}}(t',t)) \label{V_formal} . 
\end{eqnarray} 
In eq.(\ref{Lformal0}) we have defined $L_{\text{formal}}$ 
by extracting the factor $- \epsilon(t' -t)$, which is caused by the time reflection 
of eq.(\ref{time_reflection}). 
$L_{\text{formal}}$ looks like a non-translational invariant Lagrangian 
depending on $t'$, and $t$ is just a selected point in time. 
Therefore, we normally have to think of $t'$ as the time 
when using $L_{\text{formal}}$.

One may think that 
the transition amplitude of eq.(\ref{AbraAket}) can be expressed as 
\begin{eqnarray}
&&
\langle A(t) | A(t) \rangle \nonumber \\
&=&
\langle A(T_B)| e^{\frac{i}{\hbar} H^\dag (t-T_B)} 
e^{-\frac{i}{\hbar} H (t-T_A)} | A(T_A) \rangle
\nonumber \\
&=&
\int {\cal D}q {\cal D}q'  
\psi_A(q_{T_B} , T_B)^{*_q} 
e^{ \frac{i}{\hbar} S_{\text{$t$ to $T_B$}}(q)^{*_q} } 
e^{ \frac{i}{\hbar} S_{\text{$T_A$ to $t$}}(q')}  \psi_A(q'_{T_A}, T_A) 
\delta(q_t-{q'}_t) \nonumber \\
&=&
\int {\cal D}q \exp\left[ 
\frac{i}{\hbar} \int_{T_A}^{T_B} dt' 
\left\{ 
\theta(t-t') L(q) + \theta(t'-t) L(q)^{*_q} 
\right\}
\right] 
\psi_A(q_{T_B} , T_B)^{*_q} \psi_A(q_{T_A} , T_A) , \nonumber \\
&=&
\int {\cal D}q \exp\left[ 
\frac{i}{\hbar} \int_{T_A}^{T_B} dt' 
L_{\text{formal, 2}}
\right] 
\psi_A(q_{T_B} , T_B)^{*_q} \psi_A(q_{T_A} , T_A) , \label{AbraAket-futureincluded}
\end{eqnarray}
where $\theta(t)=\frac{1}{2}( \epsilon(t) + 1 )$ is a step function 
defined as 1 for $t>0$ and 0 for $t<0$, and 
$L_{\text{formal, 2}}$ is given by 
\begin{equation}
L_{\text{formal, 2}}( q(t'), \dot{q}(t') , t' -t ) 
\equiv 
\text{Re}_q L( q(t') , \dot{q} (t') )  
- i \epsilon(t' -t)  \text{Im}_q L( q(t') , \dot{q} (t') ) .  
\end{equation}
We might think that this rewriting is also good for our purpose, but 
this is not the case, since 
in eq.(\ref{AbraAket-futureincluded}) only the half of the original path, 
i.e. the path going from $T_A$ to $t$, 
is mapped onto the time interval $[T_A, T_B]$ 
over which $L_{\text{formal, 2}}$ is time-integrated.

\subsection{Momentum relation in the future-not-included theory}

Since we have found the formal Lagrangian $L_{\text{formal}}$, 
we try to obtain the momentum relation in the future-not-included theory 
by replacing $L$ with $L_{\text{formal}}$ in the method of ref.\cite{Nagao:2011is}. 
Then we obtain the formal momentum $p_{\text{formal}}(t', t)$: 
\begin{eqnarray}
p_{\text{formal}}(t', t) 
&=& \frac{\partial L_{\text{formal}}( q_{\text{formal}}(t', t), \partial_{t'} q_{\text{formal}}(t', t), t' -t)}{  \partial (\partial_{t'} q_{\text{formal}}(t', t) )} \nonumber \\
&=&m_{\text{formal}}(t' -t) \partial_{t'} q_{\text{formal}}(t', t) . \label{pformal}
\end{eqnarray}
Since $\partial_{t'} q_{\text{formal}}$ could jump up around $t'=t$, 
we take the time average of this around $t'=t$, 
expecting a finite observation time. 
Thus, the time derivative of $q(t)$, which is  given in eq.(\ref{qqformal}), is evaluated as 
\begin{eqnarray}
\frac{d}{dt} q(t) 
&\simeq& \left\{ \frac{\partial}{\partial t'} q_{\text{formal}}(t', t) \right\}|_{t'=t} \nonumber \\ 
&\simeq&
\frac{1}{2 \Delta t} \int_{t - \Delta t}^{t + \Delta t} dt' \partial_{t'} q_{\text{formal}}(t', t) \nonumber \\
&=&
\frac{1}{2 \Delta t} \int_{t - \Delta t}^{t + \Delta t} dt' 
\frac{p_{\text{formal}}(t', t) }{m_{\text{formal}}(t' -t) } \nonumber \\
&\simeq& 
\frac{1}{m_{\text{eff}}} p(t) ,  \label{momentum_relation_fni} 
\end{eqnarray}
where in the first equality we have used the relation 
\begin{equation}
\left\{ \frac{\partial}{\partial t} q_{\text{formal}}(t', t) \right\}|_{t'=t}  = 0 , 
\end{equation}
which holds because $q_{\text{formal}}(t',t)$ is independent of $t$ for $t'<t$ 
and is supposed to be smooth. 
In the second equality we have changed the expression into 
the time average of $\partial_{t'} q_{\text{formal}}$ around $t'=t$. 
In the third and fourth equalities we have used eq.(\ref{pformal}), 
and supposed that $p_{\text{formal}}$ changes very little near $t'=t$, 
and $m_{\text{eff}}$ and $p(t)$ are given by eq.(\ref{m_eff}) and 
\begin{equation}
p(t) \equiv p_{\text{formal}}(t, t) . 
\end{equation}
Thus, we have succeeded in reproducing eq.(\ref{momentum_fni}), 
the momentum relation in the future-not-included theory, 
by utilizing the method of ref.\cite{Nagao:2011is}. 
Eq.(\ref{momentum_relation_fni}) is consistent with 
eq.(\ref{ihbardeldeltqhatAA_t}), which was derived 
by analyzing the time derivative of $\langle \hat{q}_{new} \rangle^{AA}$.

Finally, we make a couple of remarks. 
If we naively average $p_{\text{formal}}$ first, 
then we might expect a relation like 
$p=m_R \dot{q}$, which is not right. 
It is  $\partial_{t'} q_{\text{formal}}$, not $p_{\text{formal}}$, 
that we should average because 
the former includes the derivative with regard to $t'$, which could jump up 
around $t'=t$.  
Similarly, it is not reasonable to take the time average of $L_{\text{formal}}$ 
because it includes $\partial_{t'} q_{\text{formal}}(t', t)$, which 
we need to average separately.

\section{Discussion}

In this paper, after reviewing 
the complex coordinate formalism\cite{Nagao:2011za},  
the method used to derive the momentum relation 
via Feynman path integral (FPI)\cite{Nagao:2011is} 
and some properties of the future-included theory studied in ref.\cite{Nagao:2012mj}, 
we provided 
the momentum relation and classical limit in the future-not-included theory, 
which are different from those in the future-included theory. 
In section~\ref{sec:review_complex_coordinate} we reviewed the 
complex coordinate formalism\cite{Nagao:2011za}, which 
is a kind of generalized bra-ket formalism so that we can properly 
deal with complex coordinate $q$ and momentum $p$. 
In section~\ref{sec:review_xi-state_argument}, 
following ref.~\cite{Nagao:2011is}, we reviewed 
the method used to derive the momentum relation by analyzing the time development 
of $\xi$-parametrized state via FPI, 
and obtained the momentum relation $p=m\dot{q}$.  
In section~\ref{sec:prop_future-included}, 
based on ref.\cite{Nagao:2012mj}, 
we saw that the quantity $\langle \hat{\cal O} \rangle^{BA}$ 
behaves as an expectation value of some operator $\hat{\cal O}$ 
in the future-included theory, and derived the momentum relation 
$\langle \hat{p}_{new} \rangle^{BA}=m\frac{d}{dt} \langle \hat{q}_{new} \rangle^{BA}$, 
which is consistent with that given in the previous section.

In section~\ref{sec:future-not-included} we studied the future-not-included theory 
and saw that the expectation value $\langle \hat{\cal O} \rangle^{AA}$
does not time-develop so cleanly compared to 
$\langle \hat{\cal O} \rangle^{BA}$ because of the presence of 
an additional anti-commutator term. 
But this anti-commutator term is a quantum fluctuation term, 
so it vanishes in the classical limit. 
Thus, we obtained the relation 
$\langle \hat{p}_{new} \rangle^{A A}
=m_{\text{eff}}
\frac{d}{d t} \langle \hat{q}_{new}  \rangle^{A A}$ 
and claimed that  $p= m_{\text{eff}} \dot{q}$ 
is the momentum relation in the future-not-included theory. 
Moreover, we argued that, 
in the future-not-included theory, 
classical theory is described not by a full action $S$ but a certain real action $S_{\text{eff}}$,  
which is not the real part of $S$.  
This is quite in contrast to the future-included theory, whose classical theory 
is described by a full action $S$. 
Furthermore, in subsection~\ref{subsec:rechoosing}, 
we offered another way to understand the time development 
of the future-not-included theory via the future-included theory. 
The above studies suggest that the method of ref.\cite{Nagao:2011is} 
for deriving the momentum relation via FPI is valid 
in the future-included theory, but not in the future-not-included theory. 
In ref.\cite{Nagao:2011is} we derived the momentum relation $p=m\dot{q}$ 
by considering a transition amplitude 
from some initial time to final time, which is similar to the transition 
amplitude in the future-included theory, but not to that 
in the future-not-included theory. 
In section~\ref{sec:reconsideration} we 
provided a way to properly apply the method of ref.\cite{Nagao:2011is} 
to the future-not-included theory by  
rewriting the transition amplitude in the future-not-included theory 
into another expression similar to the transition amplitude in 
the future-included theory, and by introducing a formal Lagrangian. 
Indeed, we explicitly showed that we 
can derive the momentum relation $p= m_{\text{eff}} \dot{q}$ 
in the future-not-included theory via this method. 
We summarized the difference between the future-included 
and future-not-included theories in Table~\ref{tab:comparison_fi_fni}.

Finally, let us seek the possibility of defining some sensible formal Hamiltonian 
in the future-not-included theory starting from the formal Lagrangian 
$L_{\text{formal}}$ in the method of ref.\cite{Nagao:2011is}, 
where we derived not only the momentum relation but 
also a Hamiltonian via the path integral. 
In section~\ref{sec:reconsideration} we provided a way to utilize the 
method by introducing $L_{\text{formal}}$. 
So if we use $L_{\text{formal}}$, we would obtain a formal Hamiltonian. 
Replacing $L$ with $L_{\text{formal}}$ 
results in replacing $m$ and $V$ with $m_{\text{formal}}(t' -t)$ 
and $V_{\text{formal}}( q(t'), t' -t )$ respectively 
in the expression of eq.(\ref{expHhat}). 
Thus, we would obtain a formal Hamiltonian, 
$\hat{H}_{\text{formal}}( t' -t ) 
= \frac{1}{2m_{\text{formal}} (t' -t) } (\hat{p}_{new})^2 + V_{\text{formal}}(\hat{q}_{new} , t' -t)$, 
where $m_{\text{formal}}(t' -t)$ and $V_{\text{formal}}( q(t'), t' -t )$ 
are given in eqs.(\ref{m_formal})(\ref{V_formal}) respectively. 
But this Hamiltonian does not have a good physical meaning.  
We can see this by trying to introduce some formal state 
$| \psi(t', t) \rangle_{\text{formal}}$, 
which time-develops according to the formal Schr\"{o}dinger equation 
$i \hbar \frac{\partial}{\partial t'} | \psi(t', t)  \rangle_{\text{formal}} 
= \hat{H}_{\text{formal}} ( t' -t ) | \psi(t', t)  \rangle_{\text{formal}}$ 
with an initial condition 
$| \psi(T_A , t) \rangle_{\text{formal}} = | A(T_A) \rangle$. 
Let us define some effective state $| \psi(t) \rangle_{\text{eff}}$ by 
$| \psi(t) \rangle_{\text{eff}} = | \psi(t , t) \rangle_{\text{formal}}$, 
where $| \psi(t', t)  \rangle_{\text{formal}}$ is supposed to be smooth 
in $t'$. 
Then taking the time average of the formal Schr\"{o}dinger equation, 
we would obtain the effective Schr\"{o}dinger equation, 
$i \hbar \left( \frac{\partial}{\partial t}| \psi(t) \rangle_{\text{eff}} 
-  \frac{\partial}{\partial t}  
| \psi(t', t)  \rangle_{\text{formal}} |_{t'=t}    \right)
= \hat{H}_{\text{eff}}  | \psi(t) \rangle_{\text{eff}}$, 
where $| \psi(t) \rangle_{\text{eff}}$ obeys an initial condition 
$| \psi(T_A) \rangle_{\text{eff}} = | A(T_A) \rangle$, 
and $\hat{H}_{\text{eff}}$ is expressed as 
$\hat{H}_{\text{eff}} 
\equiv
\frac{1}{2 \Delta t} \int_{t - \Delta t}^{t + \Delta t} dt' \hat{H}_{\text{formal}} ( t' -t )
=\frac{1}{2m_{\text{eff}}} (\hat{p}_{new})^2 + V_R(\hat{q}_{new}) 
\simeq
\hat{H}_{h}$. 
Here, in the last equality, expecting that 
this $\hat{H}_{\text{eff}}$ is to be put in some matrix elements, 
we have used the approximation $\hat{q}_{new}^\dag \simeq \hat{q}_{new}$ and 
$\hat{p}_{new}^\dag \simeq \hat{p}_{new}$ based on Theorem 1 given 
in subsection~\ref{prop_qnewpnew}. 
The effective Schr\"{o}dinger equation shows that $\hat{H}_{\text{eff}}$ is not a Hamiltonian because 
we have the second term on the left-hand side, 
though the time average of $\hat{H}_{\text{eff}}$ becomes the 
classical Hamiltonian in the future-not-included theory $H_R$, 
which is given in eq.(\ref{effective_classicalH0}). 
Therefore, $\hat{H}_{\text{formal}}$ is not a sensible Hamiltonian. 
It would be interesting if we could find some sensible formal Hamiltonian in the future, 
but practically we do not need this, because we know that 
the quantum Hamiltonian of the future-not-included theory is $\hat{H}$ 
by definition, and also we found in this paper that the 
classical theory is described by $H_{R}$.

Now that we have understood the general classical properties of 
the future-not-included complex action theory, it would be desirable 
to study the dynamics of the theory in some concrete model. 
We will work on both the future-included and future-not-included theories, 
and report some progress in the future.

\section*{Acknowledgements}

The work of K.N. was supported in part by 
Grant-in-Aid for Scientific Research (No.21740157) 
from the Ministry of Education, Culture, Sports, Science and Technology (MEXT, Japan). 
K.N. would like to thank the members and visitors of NBI for their kind hospitality, 
and M.~Fukuma, K.~Oda and S.~Sugishita for useful discussions. 
H.B.N. is grateful to NBI for allowing him to work at the institute as emeritus.



\end{document}